\documentclass[letterpaper, journal,10 pt]{IEEEtran} 

\usepackage{lettrine} 

\usepackage{balance} 

\usepackage{tikz}
\usetikzlibrary{shapes.geometric, arrows.meta, decorations.pathmorphing, decorations.markings, calc, 3d}
\usetikzlibrary{positioning, fit, backgrounds}
\tikzset{
	tangent/.style={decoration={
			markings,mark=at position #1 with {
				\coordinate (ta) at (0,0);
				\coordinate (tb) at (0.1,0);
			}
		},postaction=decorate},
	tangent/.default=0.5
}

\newcommand{\parstart}[1]{\noindent \textbf{#1.}\;}

\newcommand{\R}{\mathbb{R}}
\newcommand{\Rn}[1]{\mathbb{R}^{#1}}

\newcommand{\titlesc}[1]{\title{\Large \vspace{0.7cm} \LARGE \centering {\textsc{{#1}}}}}

\usepackage{caption} 
\usepackage{epsfig,color,amsmath,cite}
\usepackage{amsthm} 
\usepackage{amsmath}    
\usepackage[T1]{fontenc}
\usepackage[utf8]{inputenc}
\usepackage{bm}
\usepackage{epstopdf}
\usepackage{amssymb}
\usepackage{url}
\usepackage{enumitem} 
\usepackage{multirow}
\usepackage{hhline}
\usepackage{booktabs}
\usepackage{mathtools}
\usepackage{makecell}

\usepackage[linesnumbered,boxed,commentsnumbered,ruled,vlined,longend]{algorithm2e}
\SetKwProg{Init}{Initialization}{}{Proceed}
\usepackage{comment}

\DeclareMathOperator*{\subjectto}{subject\ to}

\makeatother
\DeclareMathAlphabet\mathbfcal{OMS}{cmsy}{b}{n}

\newtheorem{mydef}{Definition}



\usepackage{stackengine}

\newcommand{\mat}[1]{\boldsymbol{#1}}
\renewcommand{\vec}[1]{\boldsymbol{\mathrm{#1}}}

\newcommand{\bmat}[1]{\begin{bmatrix} #1 \end{bmatrix}}

\providecommand{\mA}{\ensuremath{\mat{A}}}
\providecommand{\mB}{\ensuremath{\mat{B}}}
\providecommand{\mC}{\ensuremath{\mat{C}}}

\providecommand{\vf}{\ensuremath{\vec{f}}}

\providecommand{\vu}{\ensuremath{\vec{u}}}

\providecommand{\vx}{\ensuremath{\vec{x}}}
\providecommand{\vy}{\ensuremath{\vec{y}}}

\newcommand{\m}{\boldsymbol}
\allowdisplaybreaks[4]
\usepackage[colorlinks = true,
linkcolor = blue,
urlcolor  = black,
citecolor = blue,
anchorcolor = blue]{hyperref}

\newcommand{\mc}[1]{\mathcal{#1}}
\newcommand{\mbb}[1]{\mathbb{#1}}
\newcommand{\mr}[1]{\mathrm{#1}}
\usepackage[framemethod=TikZ]{mdframed}
\mdfdefinestyle{MyFrame}{%
	linecolor=black,
	outerlinewidth=1.25pt,
	roundcorner=1.25pt,
	innerrightmargin=5pt,
	innerleftmargin=5pt,}
	
\usepackage{tikz}
\usetikzlibrary{matrix,positioning,decorations.pathreplacing}
\usetikzlibrary{shapes.geometric, arrows}
\usetikzlibrary{backgrounds}
\usetikzlibrary{shapes}
\usetikzlibrary{tikzmark}
\usetikzlibrary{calc}
\usetikzlibrary{arrows,shapes,positioning,shadows,trees,mindmap}
\usepackage[edges]{forest}
\usetikzlibrary{arrows.meta}
\colorlet{linecol}{black!75}
\usepackage{xkcdcolors} 

\usepackage[noabbrev]{cleveref}

\usepackage{mathtools}

\DeclarePairedDelimiter\abs{\lvert}{\rvert}%
\DeclarePairedDelimiter\norm{\lVert}{\rVert}%

\makeatletter
\let\oldabs\abs
\def\abs{\@ifstar{\oldabs}{\oldabs*}}
\let\oldnorm\norm
\def\norm{\@ifstar{\oldnorm}{\oldnorm*}}
\makeatother

\usepackage[english]{babel}
\usepackage[utf8]{inputenc}
\usepackage[super]{nth}

\usepackage{graphicx}
\usepackage{float}
\usepackage[caption = false]{subfig}

\usepackage{array}
\usepackage{threeparttable}

\usepackage[english]{babel}
\usepackage[utf8]{inputenc}
\usepackage[super]{nth}

\SetKwRepeat{Do}{do}{while}%

\everymath{\displaystyle}

\newcolumntype{E}{>{\centering\arraybackslash}m{0.48in}}
\newcolumntype{Q}{>{\centering\arraybackslash}m{2.6in}}
\newcolumntype{W}{>{\centering\arraybackslash}m{1.8in}}
\newcolumntype{K}{>{\centering\arraybackslash}m{1.6in}}

\usepackage{balance} 

\usepackage{xcolor} 
\definecolor{subsectioncolor}{RGB}{39,94,77} 

\usepackage{lettrine} 

\titlesc{Observability and Generalized Sensor Placement for Nonlinear Quality Models in Drinking Water Networks}

\author{Mohamad H. Kazm$\text{a}^{\dagger, \ast \ast}$, Salma M. Elsheri$\text{f}^{\dagger, \P}$, and Ahmad F. Tah$\text{a}^{\dagger}$ \vspace{-0.9cm}
	\thanks{$^\dagger$Department of Civil and Environmental Engineering, Vanderbilt University, Nashville, TN, USA. Emails: mohamad.h.kazma@vanderbilt.edu, salma.m.elsherif@vanderbilt.edu,  ahmad.taha@vanderbilt.edu}
	\thanks{$\P$Secondary appointment: Department of Irrigation and Hydraulics Engineering, Faculty of Engineering, Cairo University.}
	\thanks{$^{\ast \ast}$Corresponding author.}
	\thanks{This work is supported by the National Science Foundation under Grants 2151392 and 2152450.}
}
	 
\begin{document}
\setlength{\abovedisplayskip}{3.1pt}
\setlength{\belowdisplayskip}{3.1pt}
\setlength{\abovedisplayshortskip}{3.1pt}
\setlength{\belowdisplayshortskip}{3.1pt}
	
\newdimen\origiwspc%
\newdimen\origiwstr%
\origiwspc=\fontdimen2\font
\origiwstr=\fontdimen3\font

\maketitle
\thispagestyle{headings} 
\pagestyle{headings}

\markboth{Journal of Water Process Engineering, Volume 77, September 2025}{}

\begin{abstract}
This paper studies the problem of optimal placement of water quality (WQ) sensors in water distribution networks (WDNs), with a focus on chlorine transport, decay, and reaction models. Such models are traditionally used as suitable proxies for WQ. The literature on this topic is inveterate, but has a key limitation: it utilizes simplified single-species decay and reaction models that do not capture WQ transients for nonlinear, multi-species interactions. This results in sensor placements (SP) that do not account for nonlinear WQ dynamics. Furthermore, as WQ simulations are parameterized by hydraulic profiles and demand patterns, the placement of sensors are often hydraulics-dependent. This study produces a greedy algorithm that addresses the two aforementioned limitations. The algorithm is grounded in nonlinear dynamic systems and observability theory, and yields SPs that are submodular and robust to hydraulic changes. Case studies on benchmark water networks are provided. The key findings provide practical recommendations for WDN operators. 
\end{abstract}
\vspace{-0.15cm}
\begin{IEEEkeywords}
	Multi-species dynamics, water quality, monitoring, sensor placement, observability, water distribution networks
\end{IEEEkeywords}
\vspace{-0.5cm}
\section{\large Introduction and Paper Contributions}\label{sec:Into-Lit}
\subsection{Background}
\noindent \lettrine[lines=2]{S}{ensor} placement in complex infrastructure is a fundamental engineering problem. The deployment of sensors within networked systems, such as power grids, water distribution networks (WDNs), and transportation systems, enables the monitoring and safe operation of these engineered systems. From this perspective, the real-time monitoring of WDNs allows water quality (WQ) controllers to trace the evolution of disinfectants and contaminants within the network. Its aim is to regulate WQ and apply feedback control while maintaining pathogen-free water. However, to achieve effective monitoring of WQ within WDNs, the dynamical system representing the WQ model must be observable. In the context of such a dynamical system and WQ control, observability refers to the ability of water system operators to monitor the levels of contaminants and chlorine. This, in turn, allows operators to plan WQ profiles for different but typical hydraulic settings and possible, though less typical, contaminant intrusion conditions within the network.  

{While allocating sensors within the entire WDN guarantees full system observability, it is often impractical. The cost of sensors and their installation requires optimal placement at specific network locations~\cite{Rathi2014,Zeng2016}. However, observability can still be achieved through an optimal selection of a subset of sensor locations, thus ensuring system monitoring with a minimal number of sensors. The optimal sensor placement (SP) framework thereby identifies the sensor configuration required to fully characterize the WQ states in the WDN. This optimal sensor configuration set should enable the estimation of most WQ state variables. However, in practice, it is typically assumed that observability conditions are satisfied when considering the state estimation of WQ dynamics monitoring~\cite{Mankad2022}. This optimal placement of WQ sensors is considered a complex binary selection problem that becomes challenging for mid-to-large networks, particularly WQ networks with fluctuating hydraulic settings due to varying water demand patterns. Furthermore, the variability of WDN input parameters, such as nodal demands, pipe roughness coefficients, and WQ parameters, raises several challenges that must be addressed by the chosen sensor configuration~\cite{Comboul2013}. This necessitates a framework that accounts for this variability.} With that in mind, this paper aims to provide a generalized robust sensor configuration that allows the accurate depiction and estimation of multi-species WQ dynamics within WDNs under different operational and hydraulic conditions. {As compared to single-species models of WQ dynamics that depict only the decay of chlorine states within a WDN, the multi-species model captures the nonlinear interactions of chlorine with other species (e.g., microbial or chemical), whether in the bulk flow, attached to pipe walls, or arising from contamination events.}

\subsection{Relevant Literature}
The literature approaches solving the optimal geographic placement of WQ sensors using objective-based methods that assign either public health metrics related to contamination events~\cite{Mankad2022} or, more recently, network-wide observability metrics~\cite{Taha2021c} to determine the optimal sensor configuration. Comprehensive surveys on different SP optimization frameworks are presented in~\cite{Adedoja2018,Hu2018}; herein, we provide a brief review of the more recent literature.

A heuristic algorithm for solving the SP problem when sensors are not reliable is proposed in~\cite{Winter2019}. The framework utilizes a maximum set covering model with weighted edges, where reliability is defined as the probability that a sensor measurement might not be available. A dual pressure and WQ sensor selection problem is solved by considering leakage and contamination detection in WDNs~\cite{Huang2021}. The proposed method relies on a fuzzy C-means algorithm to classify the junctions via a sensitivity analysis matrix. Reference~\cite{Giudicianni2022a} introduced a new perspective that considers the placement of WQ sensors on pipes and nodes of the WDN. Reference~\cite{Zaman2022} proposes a generalized decision support framework to optimize sensor configuration for effective leakage control in sensor-deficient WDNs. This framework uses network sensitivity analysis and multi-criteria decision-making to evaluate pressure response to leaks across various hydraulic scenarios, maximizing spatial efficiency and leak detection accuracy.

The importance of considering the variability in the demand and supply schedule when making sensor placement decisions is demonstrated in~\cite{Namtirtha2023}. A comprehensive framework that further evaluates the robustness of chosen sensor configurations is developed in~\cite{Zheng2022} using an evolutionary-based optimization approach. The study shows that demand and supply have a significant influence on the robustness of the SP. {This influence is further investigated in~\cite{Giudicianni2022c}, which examines how real-time water demand fluctuations (pulsed demand) affect SP results in WDNs. The results show that accounting for pulsed demands leads to more robust and effective sensor configurations.} A dynamic graph-theoretic approach to SP in WQ networks for detecting water contamination in near real-time is proposed in~\cite{Shahra2023}. A multi-objective function (volume of contamination and time of contamination detection) is solved using an evolutionary algorithm that adapts sensor locations based on the suspected contamination source. A graph-theoretic approach to solve the optimal SP problem while maximizing the resilience of a given WQ sensor configuration in WDNs against failures is developed in~\cite{Nikolopoulos2023a}. {A framework to obtain optimal Pareto SP results for improving system resilience under a budget constraint is proposed for WDNs in~\cite{Du2025}. The results show that incorporating resilience enhances the system’s ability to handle uncertain events.} Graph centrality measures are employed in~\cite{Diao2023} to identify network coverage and time to detection, concepts related to system observability but from a network graph theory perspective.

A weighted SP optimization problem based on a discrete-time dynamical systems approach is developed in~\cite{Li2022}, considering sensor cost, information richness, and reliability. The method is applied to wastewater networks and offers observability measures for evaluating strategies related to average corrosion degree and maintenance cost. From such perspective, it is illustrated that an allocated configuration of sensors enables the detection and response to water quality issues~\cite{Li2024a}. However, the limited availability of sensors impedes the monitoring and management of WDNs. The concept of observability of a WDN, required prior to the application of estimation methods, is introduced in~\cite{Diaz2016}. An unmeasured state variable needs to be reliably estimated using available measurements; this is an essential step for the control of a WDN~\cite{Shao2023}. The impact of a chosen sensor configuration on the quality of WQ dynamics state estimation is explored in~\cite{Rajakumar2019}. The study illustrates that WQ dynamics can be estimated from a limited sensor configuration; however, the location of such sensors affects the accuracy of the state estimates. The SP problem, which considers minimizing WQ state estimation errors under time-varying WQ dynamics, is studied in~\cite{Taha2021c}. This was the first attempt to adopt a control-theoretic approach that takes into account single-species WQ observability in WDNs while solving the SP problem using a \textit{submodular} set maximization approach. Typically, sensor selection problems posed as a mixed-integer programming (MIP) formulation, as presented in~\cite{Berry2006} for WDNs, are solved using global solvers. Submodular set optimization algorithms for solving the SP problem in WDNs were first introduced in~\cite{Krause2008}, providing an efficient approach toward optimizing sensor configurations against contaminant intrusions in WDNs. 

\subsection{Research Gaps, Novelty and Contributions}
The evolution of chlorine residuals in WDNs is typically modeled with single-species decay and reaction dynamics that focus on only spatiotemporal chlorine concentrations~\cite{Elsherif2023a}. The aforementioned methods adopt the use of simplified single-species decay and reaction models. As such, the resulting sensor configurations are not robust to changes in the network's WQ dynamics that capture far more than chlorine decay. Multi-species models, however, can integrate any type of secondary reactant or aggregated reactants that react with chlorine, thereby presenting a more comprehensive model that depicts nonlinear chemical interactions with common reactants. These reactants may include naturally occurring organic matter (NOM), ammonia, and nitrite, all of which are known to consume chlorine, affect its residual levels, and result in the formation of disinfection by-products (DBPs), which are regulated due to their potential health risks~\cite{Wang2013}. Thus, as compared to single-species models, multi-species WQ models allow for the allocation of sensors while accurately accounting for WQ species reactivity under varying operational conditions. To that end, in this work, we introduce a robust observability-based multi-species WQ sensor placement  framework. The underlying WQ model is based on a multi-species reaction dynamics representation; it enables {nonlinear} contaminant reactivity modeling. Furthermore, we leverage the submodular properties of the constructed nonlinear observability measures to solve for the optimal WQ sensor configuration problem. Under such formulation, greedy algorithms are employed to solve the combinatorial set optimization problem. Accordingly, as compared to the approach developed in~\cite{Taha2021c}, the proposed SP framework offers the following: \textit{(i)} a generalized robust solution to fluctuations in water demand patterns and varying WQ conditions; \textit{(ii)} a scalable algorithm that enables its applicability to large-scale networks, resulting from the submodularity of the proposed WQ sensor placement problem; and \textit{(iii)} a dynamical model that considers the nonlinear multi-species WQ dynamics.

{The paper's main contributions follow.}
\begin{itemize}[leftmargin=*]
	\item We introduce a control-theoretic approach for quantifying the observability of the nonlinear multi-species WQ dynamics in WDNs. 
	The observability framework is based on a moving horizon formulation that accounts for varying WQ dynamics, depicting the evolution of spatiotemporal multi-species dynamics. The proposed nonlinear observability metric is used to quantify the observability of  WQ states and is related to the estimation of WQ concentrations resulting from a given sensor configuration. The proposed observability formulation extends the framework introduced in our previous work~\cite{Kazma2023f} for general nonlinear systems. That is, we extend the formulation to nonlinear difference equation models of WQ in drinking water networks.
	\item We provide robust observability measures that ensure the submodularity and modularity of the sensor selection objective function are retained. In particular, we leverage the modularity and submodularity constructed observability measures to perform the SP. Based on such measures we attain an optimal sensor configuration that is robust towards varying demand patterns and WQ conditions/parameters. Furthermore, under this formulation, greedy algorithms are employed to solve the combinatorial set optimization problem, rendering the SP problem scalable for larger networks. As compared to the previous results in~\cite{Taha2021c}, solving the SP problem under the proposed framework yields a single sensor configuration that is optimal under varying nonlinear hydraulic and water quality conditions.
	\item A comprehensive case study is provided on benchmark water networks with different scales of water networks, number of allocated sensors, water demand profiles, and WQ conditions. The sensor placement framework is solved by considering several system observability measures. This offers different observations that WDN system operators can consider, such as varying hydraulic profiles arising from daily water demand patterns, the interactions of different WQ reactants and contamination intrusion.
\end{itemize}

\parstart{Notation} Let $\mbb{N}$, $\R$, $\Rn{n}$, and $\Rn{p\times q}$ denote the set of natural numbers, real numbers, real-valued row vectors with size of $n$, and $p$-by-$q$ real matrices respectively. The cardinality of a set $\mc{N}$ is denoted by $|\mc{N}|$. The symbol $\otimes$ denotes the Kronecker product. The operators $\mr{det}(\m{A})$ returns the determinant of matrix $\m{A}$,  $\mr{trace}(\m{A})$ returns the trace of matrix $\m{A}$. The operator $\{\vx_{i}\}_{i=0}^{N}  \in \Rn{Nn}$ constructs a column vector that concatenates vectors $\vx_i \in \Rn{n}$ for all $i \in \{0, 1,\ldots, N\}$. The operator $ \mr{diag}{\{\gamma_j\}}_{{j}=1}^{n}\in \Rn{n\times n}$ constructs a diagonal matrix of vector $\gamma\in \mbb{R}^{n}$. For any two matrices $\m{A}$ and $\m{B} \in \Rn{n\times n}$, the notation $\{\m{A}, \m{B}\}$ represents $\left[\m{A}^{\top} \m{B}^{\top}\right]^{\top}$.

\parstart{Paper Organization} The paper is organized as follows:~Section~\ref{sec:mult-spec-dynamics} introduces the WQ dynamics model.~Section~\ref{sec:Obs-WQD} provides preliminaries on observability quantification and introduces the proposed WQ observability measures.~In Section~\ref{sec:SP-WQ}, we introduce the proposed optimal SP framework. The numerical case studies are presented in Section~\ref{sec:casestudy}, and Section~\ref{sec:summary} provides a conclusion to the paper while addressing the paper's limitations and future work. 
\vspace{-0.3cm}
\section{Multi-species WQ Dynamics Model}\label{sec:mult-spec-dynamics}
We model the WDN by a directed graph $\mc{G} = (\mc{N},\mc{L})$. The set $\mc{N}$ defines the nodes and is partitioned as $\mc{N} = \mc{J} \cup \mc{T} \cup \mc{R}$ where sets $\mc{J}$, $\mc{T}$, and $\mc{R}$ are collections of junctions, tanks, and reservoirs. The directed nature of the graph is due to the flow directions within the network. Let $\mc{L} \subseteq \mc{N} \times \mc{N}$ be the set of links, and define the partition $\mc{L} = \mc{P} \cup \mc{M} \cup \mc{V}$, where sets $\mc{P}$, $\mc{M}$, and $\mc{V}$ represent the collection of pipes, pumps, and valves. Total number of states is $n_x=n_L+n_N$, where $n_\mr{L}$ and $n_\mr{N}$ are numbers of links and nodes. The states depict the time-evolution of the concentrations of chlorine and its reactant within the WDN components.
The number of reservoirs, junctions, tanks, pumps, valves, and pipes are $n_\mr{R}, n_\mr{J}, n_\mr{TK}, n_\mr{M}, n_\mr{V},$ and $n_\mr{P}$. Each pipe $i$ with length $L_i$ is spatially discretized and split into $s_{L_i}$ segments. Hence, the total number of links states can be  expressed as $n_\mr{L}=n_\mr{M}+n_\mr{V}+ \sum_{i=1}^{n_\mr{P}} s_{L_i}$ and the number of nodes as $n_\mr{N}=n_\mr{R}+n_\mr{J}+n_\mr{TK}$.

The dynamics involving multiple species are described through nonlinear difference equations (NDEs), capturing the intricate interactions between chlorine and another fictitious reactant as they evolve across the entire network. The fictitious reactant can depict any real-system reactant present within a WDN under study. It may represent a single reactant, such as NOM, ammonia, or nitrite, or an aggregate of all reactants within the WDN. The use of a fictitious reactant within this manuscript allows for a generalized SP framework for multi-species WDNs. The NDEs are structured according to the subsequent state-space representation, allowing for the depiction of chemical evolution, injections at booster stations, and measurements from sensors. Equations~\eqref{equ:NDE1} and~\eqref{equ:NDE2} describe the NDE state-space dynamics and measurement equations governing the concentration evolution of interacting species across the WDN;
\begin{subequations}\vspace{0.2cm}\label{equ:NDEs}
	\begin{align}
		\begin{split}
		\vspace{-0.3cm}	\underbrace{	\begin{bmatrix}
					\vx_1(k+1) \\ \vx_2(k+1)
			\end{bmatrix}}_{{\vx}(k+1)} & = 
			\underbrace{\begingroup 
				\setlength\arraycolsep{2pt}	\begin{bmatrix}
					\mA_{11}(k) & 0 \\ 0 & \mA_{22}(k)
				\end{bmatrix} \endgroup}_{{\mA}(k)}
			\underbrace{ \begin{bmatrix}
					\vx_1(k) \\ \vx_2(k)
			\end{bmatrix}}_{{\vx}(k)} \\ & +  \underbrace{\begingroup 
				\setlength\arraycolsep{2pt}	\begin{bmatrix}
					\mB_{11}(k) & 0 \\ 0 & \mB_{22}(k)
				\end{bmatrix} \endgroup}_{{\mB}(k)} 
			\underbrace{\begin{bmatrix}
					\vu_1(k) \\ \vu_2(k)
			\end{bmatrix}}_{{\vu}(k)} + \vf(\vx_1,\vx_2,k), \label{equ:NDE1}
		\end{split}
		\\ \begin{split}
			\vspace{-0.3cm}	\underbrace{	\begin{bmatrix}
					\vy_1(k) \\ \vy_2(k)
			\end{bmatrix}}_{{\vy}(k)} & = 
			\underbrace{	\begin{bmatrix}	\mC_{11}(k) & 0 \\
					0 & \mC_{22}(k)
			\end{bmatrix}}_{{\mC}(k)} 
			\underbrace{ \begin{bmatrix}
					\vx_1(k) \\ \vx_2(k)
			\end{bmatrix}}_{\vx(k)}, \label{equ:NDE2}
		\end{split}
	\end{align}
\end{subequations}
where variable $k \in \mathbb{N}$ is the discrete time-step; vectors $\vx_1(k) \in \mathbb{R}^{n_{x_1}}$ and $\vx_2(k) \in \mathbb{R}^{n_{x_2}}$ depict  the concentrations of chlorine and the other fictitious reactant (two species model) in the entire network, such that  $\vx(k)= \{\vx_1(k), \vx_2(k)\} \in \mathbb{R}^{n_x}$ concatenates the vectors $\vx_1(k)$ and $\vx_2(k)$. Vector $\vu_1(k) \in \mathbb{R}^{n_{u_1}}$ represents the dosages of injected chlorine; vector $\vu_2(k) \in \mathbb{R}^{n_{u_2}}$ accounts for planned or unplanned injection of the fictitious component; vector $\vf(\vx_1,\vx_2,k)$ encapsulates the nonlinear part of the equations representing the mutual nonlinear reaction between the two chemicals; vector $\vy_1(k) \in \mathbb{R}^{n_{y_1}}$ denotes the sensor measurements of chlorine concentrations at specific locations in the network while $\vy_2(k) \in \mathbb{R}^{n_{y_2}}$ captures the fictitious reactant measurements by sensors in the network if they exist. The vector $\vy(k)= \{\vy_1(k), \vy_2(k)\} \in \mathbb{R}^{n_y}$ concatenates the vectors $\vy_1(k)$ and $\vy_2(k)$. Note that the state-space matrices $\{\mA, \mB, \mC\}_{\bullet}$ are all time-varying matrices that depend on the network layout, hydraulic parameters, decay and reaction rate coefficients for the disinfectant, and booster stations and sensors locations. It is customary to assume that these matrices evolve at a slower pace than the states $\vx(k)$ and control inputs $\vu(k)=\{\vu_1(k), \vu_2(k)\} \in \mathbb{R}^{n_u}$. Note that, matrix $\mC(k)$ determines the state-output mapping and encodes the sensor locations used to measure chlorine and the fictitious species. We parameterize matrix $\mC(k)$ to include sensor locations in Section~\ref{sec:nonlinear-obs}.

We simulate the concentration evolution throughout network components using the principles of mass conservation, transport, decay, and reaction models for these substances (utilizing a one-dimensional advection-reaction model). We note that the notations and terminology used herein are borrowed from~\cite{Elsherif2023a}. A comprehensive derivation of the models for each component is described in~\cite{Elsherif2023a}. Furthermore, these models are validated against EPANET~\cite{rossman} and its extension, EPANET-MSX~\cite{Shang2011}, which is a WQ multi-species simulation tool. For brevity, we present only the final formulations for each component in Tab. \ref{tab:HydModel} (Appendix~\ref{apndx:hydraulic_model}), allowing readers to access these details. Note that the equations listed in Tab. \ref{tab:HydModel} are for chlorine concentrations $c(k)$, however, they also apply for the fictitious reactant concentrations $\tilde{c}(k)$ at the network's components. The state-space representation \eqref{equ:NDEs} concatenates the equations in Tab.~\ref{tab:HydModel} such that the following denote the state-space vectors;
\begin{eqnarray*}
	\hspace{-0.2cm}& \vx_1(k)  := \{ c_i^\mr{N}(k), c_i^\mr{L}(k) \}, \;\;\;
	\vx_2(k) := \{ \tilde{c}_i^\mr{N}(k), \tilde{c}_i^\mr{L}(k) \},	\\	
	\hspace{-0.2cm}& \vu_1(k) := \{c^\mr{B_\mr{J}}_i(k),  c^\mr{B_\mr{TK}}_i(k)\}, \;\;\; \vu_2(k) := \{\tilde{c}^\mr{B_\mr{J}}_i(k),\tilde{c}^\mr{B_\mr{TK}}_i(k)\}.
\end{eqnarray*}

That being said, the representation in~\eqref{equ:NDEs} can be considered a generalized multi-species model that enables the time-evolution of a fictitious reactant as $\tilde{c}(k)$ (representing single or multiple chemicals that can react with chlorine) and chlorine as $c(k)$.
Furthermore, the time-space discretization of pipe $i$ is based on the upwind Eulerian discretization scheme~\cite{Basha2007}; see~\cite{Elsherif2023a} for additional information regarding the different discretization methods. The discretization method considered in this manuscript accurately represents the physical processes governed by advection-reaction dynamics, which is essential for accurate WQ modeling and the optimal placement of WQ sensors.
\vspace{-0.3cm}
\section{Observability for Nonlinear WQ Dynamics}\label{sec:Obs-WQD}
The aforementioned section presented a brief summary of the multi-species WQ model and its state-space representation. In this sequel, we introduce the notions of linear and nonlinear observability from a dynamical systems perspective and provide the multi-species WQ observability measures.
\vspace{-0.3cm}
\subsection{Linear Observability Review}\label{sec:Obs}
For a more self-contained paper, we introduce the concept of observability for linear systems and then present its extension to the nonlinear form of a WDN. Consider the NDEs introduced in~\eqref{equ:NDEs} rewritten as a linear time-varying system by eliminating the non-linearity $\vf(\vx_1,\vx_2,k)$. Note that the non-linearity part models the multi-species interactions. The linear system along with its measurement model can be succinctly written as follows
\begin{subequations}\label{eq:LNDEs}
	\begin{align}
		{{\vx}(k+1)} & = 
		{{\mA}(k)} {{\vx}(k)} + {{\mB}(k)}{{\vu}(k)}, \label{eq:LNDE1}\\
		{{\vy}(k)} & = 
			{{\mC}(k)}{\vx(k)}, \label{eq:LNDE2}
	\end{align}
\end{subequations}
where state-space matrices $\{\mA, \mB, \mC\}$ represent the same matrices introduced in Section~\ref{sec:mult-spec-dynamics}. The system~\eqref{eq:LNDEs} is said to be observable for observation horizon $N_s$ if and only if the observability matrix $\mc{O}(\vx(0)):=\mc{O}\in \mathbb{R}^{N_{s}n_{y} \times n_{x}}$, defined as~\eqref{eq:Obs_matrix}, is full column rank~\cite{Kalman1962}:
\begin{equation}\label{eq:Obs_matrix}
	\mc{O}\hspace{-0.01cm} :=\hspace{-0.01cm}\left\{\m{C}(k), \m{C}(k) \mA(k), \ldots, \m{C}(k) \mA(k)^{N_{s}-1}\right\}.
\end{equation}

Observability that is conditioned on the full rankness of the matrix $\mc{O}$ can be defined as the ability to derive the WQ system states $\vx(k)$ from the knowledge of measurements $\vy(k)$ for a finite time-horizon $N_{s}$. In terms of WQ modeling, this translates into the ability to depict the evolution of all the chlorine and fictitious reactant concentrations within the WDN from measuring the output of the system. Note that the system output, i.e., the measurements, need not measure all the concentrations within the system in order to reconstruct the full system states. Meaning that, we can employ a limited number of sensors within a  WDN and still achieve full observability of the WQ states. This observability rank condition discussed above is a qualitative metric that indicates the ability to infer states of a dynamical system by simply measuring its output. However, this is not indicative of how observable the dynamical system is. Meaning that, from this assessment we cannot identify the best sensor configuration that enables the best WQ dynamics estimation. As such, to quantify WQ dynamics~\eqref{eq:LNDEs} observability within a WDN, we introduce the concept of observability Gramian $\m{W}_{\m{o}}  \in \mathbb{R}^{n_x \times n_x}$ defined as follows
\begin{equation}\label{eq:LinObsGram}
	\m{W}_{\m{o}}:=\sum_{k=0}^{N_s-1}\m{A}(k)^{\top} \m{C}(k)^{\top} \m{C} (k)\m{A}(k) = \mc{O}^{\top} \mc{O},
\end{equation}
where the Gramian $\boldsymbol{W}_{\m{o}}$ is a positive semidefinite matrix that provides a volumetric energy-related quantification measure of observability. As such, matrix $\m{W}_{\m{o}}$ is non-singular if the system is observable over horizon $N_s$, otherwise, it is not observable.
There exist several observability measures and metrics that offer a quantification metric that extends the observability rank condition of matrix~\eqref{eq:Obs_matrix}. Such measures are expressed on the basis of the $\mr{rank}(\m{W}_{o})$, smallest eigenvalue $\lambda_{\min}(\m{W}_{o})$, $\mr{trace}(\m{W}_{o})$, and log-determinant  $\mr{log}\mr{det}(\m{W}_{o})$ of the observability Gramian~\eqref{eq:LinObsGram}; see \cite{Pasqualetti2014,Summers2016} and references therein for additional information regarding the control-theoretic observability measures.

\begin{figure}[t]
	\centering
	\subfloat{
		\resizebox{0.48\textwidth}{!}{
			\begin{tikzpicture}
				\definecolor{teal}{RGB}{0,128,128}
				\definecolor{brown}{RGB}{165,42,42}
				
				\draw[thick, black] (0, 0) -- (12, 0);
				
				\foreach \i in {0, 4, 8, 12} {
					\fill[teal] (\i, 0) circle (0.3);
				}
				\foreach \i in {1, 2, 3, 5, 6, 7, 9, 10, 11} {
					\fill[brown] (\i, 0) circle (0.2);
				}
				
				\draw[teal, thick] (0, 0.5) -- (0, 1) -- (4, 1) -- (4, 0.5);
				\draw[teal, thick] (4, 0.5) -- (4, 1) -- (8, 1) -- (8, 0.5);
				\draw[teal, thick] (8, 0.5) -- (8, 1) -- (12, 1) -- (12, 0.5);
				
				\foreach \i in {0, 4, 8} {
					\draw[brown, thick] (\i, -0.5) -- (\i, -1) -- (\i + 1, -1) -- (\i + 1, -0.5);
					\draw[brown, thick] (\i + 1, -0.5) -- (\i + 1, -1) -- (\i + 2, -1) -- (\i + 2, -0.5);
					\draw[brown, thick] (\i + 2, -0.5) -- (\i + 2, -1) -- (\i + 3, -1) -- (\i + 3, -0.5);
				}
				\foreach \i in {12} {
					\draw[brown, thick] (\i, -0.5) -- (\i, -1);}
				
				\draw[brown, thick] (3, -1) -- (9, -1);
				\draw[brown, thick] (9, -1) -- (11, -1);
				\draw[brown, thick] (11, -1) -- (12, -1);
				
				\node at (6, 1.3) {$\Delta t_H$};
				\node at (6, -1.3) {$\Delta t_{WQ}$};
	\end{tikzpicture}}}
	\vspace{-0.1cm}
	\caption{Time-steps for both the hydraulic $(\Delta t_H)$ and WQ $(\Delta t_{WQ})$ dynamics.}\label{fig:timestep}
	\vspace{-0.4cm}
\end{figure}
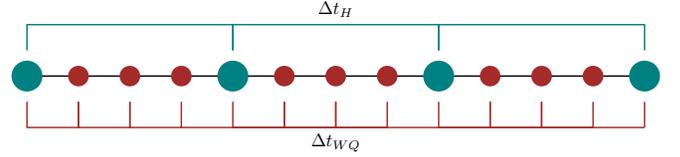

Note that, the hydraulics within the WQ time-step $\Delta t_{\mr{WQ}}$ is constant and only changes according to the hydraulic time-step $\Delta t_{\mr{H}}$. That is, the WQ time-step, denoted as $\Delta t_{\mr{WQ}}$, is slower and is updated within the hydraulic time-step $\Delta t_{\mr{H}}$ (as illustrated in Fig. \ref{fig:timestep}). As such, the WQ state-space matrices are time-variant throughout the observation window, $N_s:= \tfrac{T_s}{\Delta t_{\mr{WQ}}}$, due to changes in hydraulic dynamics. Variable $T_s$ denotes the simulation period.

This classical linear observability notion is considered in~\cite{Taha2021c} for observability-based SP under a single-species WQ model in drinking water networks.
That being said, the use of a linear model of the WQ dynamics, as in~\cite{Taha2021c}, allows for WQ sensor placement applications that are not robust to changes in the network’s nonlinear WQ dynamics under a varying hydraulic profile. Nonetheless, the use of nonlinear multi-species models (WQ model~\eqref{equ:NDEs}) necessitates observability quantification methods that capture the interactions of several reactive species while providing robustness toward fluctuating water demand patterns. The reason is that the linear observability Gramian~\eqref{eq:LinObsGram} is obtained as a unique solution of the Lyapunov equation. This equation is fundamental to control theory and is used to assess linear system stability and observability. However, it does not account for the varying nature of nonlinear WQ dynamics.

\vspace{-0.3cm}
\subsection{Observability Measures for Multi-Species WQ Dynamics}\label{sec:nonlinear-obs}
To that end, we introduce the notion of observability for nonlinear dynamical systems, specifically systems such as the multi-species WQ model. Observability of the nonlinear dynamical system is then defined as the ability to identify the initial states $\vx(0):=\vx_{0}\in \mc{X}_0$ for $t >0$. We refer to the following definitions for observability of nonlinear systems introduced in~\cite{Hermann1977}. Distinguishability can be defined as the ability to distinguish any two points  $\vx^1_0$ and $\vx^2_0 \in \mathbfcal{X}_0$ from the states at time $t>0$, that is, the system is indistinguishable if and only if for any $\vx(k):=\vx_{k} \in \mathbfcal{X}$, we have $\vx^1_k= \vx^2_k$. The study in~\cite{Hanba2017} shows that an observation window of finite length $N>0$ exists given the realization of distinguishability. Such sequence of measurements uniquely determines the initial state $\vx_0$ on the compact set $\mathbfcal{X}_{0}$. The existence of such finite-observation window defines the \textit{uniform observability} (Definition~\ref{def:uniformobs}) criterion for discrete-time models of nonlinear systems.
\vspace{-0.15cm}
\begin{mydef}[uniform observability~\cite{Hanba2009}]\label{def:uniformobs}
	Let $N>0$ denote the finite observability window, then the system is uniformly observable over compact set $\mathbfcal{X}$ with output sequence,
	\begin{equation}\label{eq:UniformObs}
		\m{\xi}(\vx_0) := \big \{\vy_{i} \big \}_{i=0}^{N_s-1}
		\in \mathbb{R}^{N_sn_y},
	\end{equation} that is injective with respect to $\vx_0 \in \mathbfcal{X}_0$.
\end{mydef}

The injectivity of the output sequence $\m{\xi}(\vx_0)$ implies that the Jacobian of the output sequence $\m{\xi}(\vx_0)$ is full rank, i.e., $\mr{rank}\left(\tfrac{\partial\m{\xi}(\vx_0)}{\partial \vx_0}\right) = n_x\; \forall \; \vx_{0} \in \mathbfcal{X}_0.$ Indeed, there exist several nonlinear observability methods that can be utilized for the quantification of nonlinear WQ observability. One approach is to compute the empirical observability Gramian of the system~\cite{Lall1999,Qi2015}. This is related to the linear observability Gramian~\eqref{eq:LinObsGram}, however, there is no clear methodology for scaling the internal states and output measurements so that the Gramian's eigenvalues capture the local variations in states~\cite{Krener2009}. Lie derivatives can be utilized however, such an approach is typically not considered for nonlinear observability under the context of sensor selection since the resulting observability measure is a rank condition that is qualitative in nature~\cite{Krener1983, Krener2009}. To that end, we refer to the quantification framework introduced in~\cite{Haber2018}. The approach adopts a moving horizon formulation that is considered intrinsically robust to measurement noise~\cite{Alessandri2008} and is more scalable as compared to empirical methods such as the Empirical Gramian. We then extend this observability quantification framework for sensor selection, offering state-robust solutions under different operating conditions, as proposed in our previous work~\cite{Kazma2023f}. 

The equivalent of nonlinear multi-species WQ dynamics~\eqref{equ:NDEs} can be compactly written, along with the parameterized measurement equation, in the following form
\begin{subequations}\label{eq:DT_dynamics}
	\begin{align}
		\vx_{k+1} &= \tilde{\m f} (\vx_{k}), \label{eq:DT_dynamics_compact} \\
		\vy_k &= \widetilde{\mC}\vx_{k} =\m{\Gamma}{\m{C}} \vx_{k}, \label{eq:DT_measurement_model}
	\end{align}
\end{subequations}
where the function $\tilde{\m f}(\cdot)$ in \eqref{eq:DT_dynamics_compact} concatenates the linear and nonlinear dynamics on the right hand side of~\eqref{equ:NDEs}. The matrix $\m\Gamma := \mr{diag}{\{\gamma_j\}}_{{j}=1}^{n_y}\in \mbb{R}^{n_y\times n_y}$ determines the configuration of the sensors within the WDN. That is, $\gamma_j = 1$ if node $j$ is equipped with a WQ sensor, and $\gamma_j = 0$ otherwise. Furthermore, we define the WQ sensor parameterization vector $\m \gamma$ representing the sensor selection as $\m \gamma:=\{{\gamma_j}\}_{j=1}^{n_y}\hspace{-0.05cm}$. The parameterized measurement mapping function $\widetilde{\mC}$ is obtained by compressing the zero rows of $\m\Gamma \m C$. 

Based on the definition of nonlinear observability (Definition~\ref{def:uniformobs}), we present the following nonlinear observability matrix that quantifies the observability of multi-species WQ dynamics within a WDN;
\begin{equation}\label{eq:obs_gramian_x0}
		\m W(\m \gamma, \vx_{0}) := \m {J}^{\top}(\m\gamma,\vx_{0})\m {J}(\m \gamma,\vx_{0}),
\end{equation}
where $\m{W}(\cdot): \Rn{n_x}\rightarrow\Rn{n_x\times n_x}$ defines the observability matrix of the nonlinear multi-species WQ dynamics. The Jacobian matrix $\m {J}(\cdot)\in\mbb{R}^{N_sn_y\times n_x}$ of output sequence $\m{\xi}(\vx_0)$  can be written as
\begin{equation}\label{eq:obs_jacobian_x0}
	\m {J}(\m \gamma,\m x_0) := \frac{\partial\m{\xi}(\vx_0)}{\partial \vx_0}
	= \left\{\frac{\partial\vy_i}{\partial \vx_0}\right\}_{{i}=0}^{N_s-1}.
\end{equation}

Taking into account the parameterized measurement equation~\eqref{eq:DT_measurement_model}, the Jacobian matrix $\m {J}(\cdot)$ in \eqref{eq:obs_jacobian_x0} for a measurement model configuration defined by $\m\gamma$ around a specific initial state $\vx_{0}$ can be written as
\begin{equation}\label{eq:obs_jacobian_x0_param}
	\m {J}(\m \gamma,\vx_0) = \bmat{\m I \otimes \m \Gamma \m C} \times \left\{\frac{\partial\vx_i}{\partial \vx_0}\right\}_{{i}=0}^{N_s-1},
\end{equation}
where matrix $\m{I} \in \Rn{n_x\times n_x}$ is an identity matrix. It is important to note that the computation of $\tfrac{\partial \vx_{i}}{\partial \vx_{0}}$ in \eqref{eq:obs_jacobian_x0_param} requires the knowledge of $\m x_i \in \mathbfcal{X}$ for all $i$. That being said, the partial derivative of output sequence $\m{\xi}_{i}(\vx_0)$  for $i \in \{0,1,\dots,N_s-1\}$ can be computed using the chain rule and is given as 
\begin{equation}\label{eq:partial_xi_0}
	 \frac{\partial\m{\xi}_{i}(\vx_0)}{\partial \vx_0}
	:= \widetilde{\mC}\frac{\partial \vx_{i}}{\partial \vx_{0}}
	= \widetilde{\mC}\prod_{i=1}^{N_s-1} \dfrac{\partial \vx_{i}}{\partial \vx_{i-1}}.
\end{equation}

Motivated by the fact that matrix $\m{W}$ is closely related to the observability Gramian~\eqref{eq:LinObsGram} of the linearized dynamics and, particularly, to the empirical observability Gramian~\cite{Lall1999,Krener2009}, we will refer to this matrix as the observability Gramian of system~\eqref{equ:NDE1}-\eqref{equ:NDE2}. Notice that the Gramian matrix~\eqref{eq:obs_gramian_x0} contains the matrix $\widetilde{\mC}$, which is a function of the vector $\m \gamma$. The observability matrix~\eqref{eq:obs_gramian_x0} is essential for the observability analysis of the nonlinearities captured within the multi-species WQ dynamics; it enables the observation of WQ species interactions. Specifically, the spectral properties of matrix $\m{W}$ allow for quantifying the ability to observe and infer the multi-species WQ dynamics within a WDN. As mentioned earlier, there exist a plethora of spectral measures that provide a scalar quantification of the linear observability Gramian~\eqref{eq:LinObsGram}. The same spectral measures can be applied to the nonlinear observability Gramian~\eqref{eq:Obs_matrix} in order to provide quantitative measures for system observability.

In the context of WQ monitoring, the aforementioned measures can have several interpretations that provide different insights to WDN operators. The $\mr{rank}$ measure, which quantifies the observable WQ dynamics space, can infer the extent of WQ monitoring coverage. That is, having a full rank observability matrix allows for the reconstruction and monitoring of all WQ dynamics under a certain sensor configuration within the WDN. The $\mr{trace}$ and $\mr{logdet}$ average and volumetric observability energy in all directions of the WQ dynamics. As such, maximizing such measures increases the ability and accuracy of the estimation and monitoring of WQ dynamics. To identify locations that have minimal observability the $\lambda_{\mr{min}}$ measure can be used. The above discussions offer measures that quantify the observability of multi-species WQ model~\eqref{equ:NDEs}. A relevant aspect of such observability notions is the ability to infer or reconstruct WQ dynamics from measurement outputs $\vy_{k}$ under a certain sensor configuration. The subsequent section details the methodology for estimating the WQ dynamics.

It is considered impossible to accurately reconstruct all WQ multi-species concentrations (including chlorine states) unless WQ sensors are readily available and widely installed at each junction within the WDN. This is due to the relationship between system observability and the ability to estimate system states. Nonetheless, to reconstruct the states for a given sensor configuration based on the presented nonlinear observability definitions, the WQ state estimation can be related to the concept of an open-loop lifted observer framework.

Based on Definition~\ref{def:uniformobs}, the multi-species WQ model ~\eqref{eq:DT_dynamics_compact} along with the measurement model \eqref{eq:DT_measurement_model} is said to be uniformly observable in $\mathbfcal{X}$ ($\mathbfcal{X}$ is the subset representing operating regions of \eqref{equ:NDEs} for a given hydraulic setting) if there exists a finite horizon $N_s \in \mbb{N}$ such that the relation $\tilde{\vy} = \m{g}\left(\hat{\vx}_{0}\right)$ is injective (one-to-one) with respect to $\vx_{0}\in \mathbfcal{X}_{0}$ for any given set of measured outputs ${\vy}_{k} = \widetilde{\mC} \vx_{k} $. We note here that the vector $\hat{\vx}\in \Rn{n_x}$ represents the reconstructed WQ states. The vector $\tilde{\vy} \in \Rn{N_sn_y}$ represents the lifted vector that is constructed as $\tilde{\vy} = \{{\vy}_{i}\}_{i=0}^{N_s-1} $. As such, we can define the function $\m{h}(\hat{\vx}_{0}):\Rn{n_x}\rightarrow \Rn{N_s n_y}$ as
\begin{equation}\label{eq:initial_state_est_fun}
	\m{h} := \tilde{\vy} - \m{g}(\hat{\vx}_{0})=  \{\vy_{i}\}_{i=0}^{N_s-1} - \{\widetilde{\mC}\hat{\vx}_{i}\}_{i=0}^{N_s-1}.
\end{equation}
From \eqref{eq:DT_dynamics_compact}, we observe that $\m{g}_{i}$ is a function of only of the initial state $\vx_0$ due to the fact that $\vx_{i}$ is a recursive function of $\vx_{0}$ for each $i$. The initial states of the WQ dynamics depend on the hydraulic conditions defined by the water demand patterns. That means, in practice the actual WQ states are unknown, then for a fixed sensor configuration within the WDN, we can estimate $\vx_{0}$ by solving a nonlinear state estimation optimization problem.
\vspace{-0.3cm}
\section{Robust SP for Multi-species WQ Dynamics}\label{sec:SP-WQ}
In this section, we introduce the robust optimal WQ sensor placement (SP) formulation for WDNs. First, we present the SP problem formulation and derive the nonlinear observability measures. Then, we propose the robust observability-based objective functions for the SP problem, and finally, we outline the algorithm used to solve the problem.
\vspace{-0.2cm}
\subsection{\small Problem Formulation and Observability Measures Derivation}\label{sec:Fomulation_SP_MSWQ}
We present the concept of set optimization as a combinatorial optimization approach with lower computational complexity. The method is based on defining a state-averaged observability measure and applying a greedy algorithm to efficiently solve the problem. The justification for using the greedy algorithm is established by demonstrating that the introduced set function measures retain the properties of set functions—namely, modularity and submodularity. The following definition characterizes modular and submodular set functions \cite{Lovasz1983,Kazma2023f}.

A set function $\mc{O}: 2^{\mc{V}}\rightarrow \R$ is modular if and only if for any $\mc{S}\subseteq\mc{N}$ and weight function $w:\mc{N}\rightarrow \R$, we have 	$\mc{O}(\mc{S}) = w(\emptyset) + \sum_{s\in\mc{S}} w(s)$. The set function is submodular if and only if for any $\mc{A},\mc{B}\subseteq\mc{N}$ given that $\mc{A}\subseteq\mc{B}$ and $s\notin\mc{B}$, we have $\mc{O}(\mc{A}\cup\{s\}) - \mc{O}(\mc{A})\geq 	\mc{O}(\mc{B}\cup\{s\}) - \mc{O}(\mc{B})$. Intuitively, for any submodular function, the addition of an element $s$ to a smaller subset $\mc{A}$ yields a greater gain compared to adding the same element to a bigger subset $\mc{B}$. This notion is known as the \textit{diminishing returns property} \cite{Summers2016}. Furthermore, the notion of monotone increasing functions is also important to achieve a scalable SP formulation; it is defined as follows. Set function $\mc{O}: 2^{\mc{N}}\rightarrow \mbb{R}$ is called monotone increasing if $\mc{A},\mc{B}\subseteq\mc{N}$, $\mc{A}\subseteq\mc{B}$ implies $\mc{O}(\mc{B})\geq 	\mc{O}(\mc{A})$. As discussed in Section~\ref{sec:nonlinear-obs}, the observability measures utilized in the context of SP have been shown to be either submodular or modular~\cite{Kazma2023f}.

We now define a set function $\mc{O}{(\mc{S})}: 2^{\mc{N}}\rightarrow \mbb{R}$ with $\mc{N} = { i\in\mbb{N},|, 0 < i \leq n_\mr{N}}$, where the set $\mc{N} = \mc{J} \cup \mc{T} \cup \mc{R}$ represents the possible set of sensor locations such that $n_y = n_\mr{N}$. Note that the set $\mc{N}$ contains the possible locations within a WDN at various junctions, tanks, and reservoirs (nodes of the WDN). The set $\mc{S}$ represents a set of sensor combinations such that $\mc{S}\subseteq \mc{N}$. In this context, the specific geographic placement and locations of these $|\mc{S}|=r$ sensors are encoded in matrix $\widetilde{\mC}$ through binary parameterization $\m{\gamma}$. Given these definitions, the SP set optimization problem for a multi-species WQ drinking water network can be written as
\begin{equation}\label{eq:OSP}
	{(\mathbf{P1})}\;\;	\mc{O}^*_{{\mc{S}}} 
	:= \max_{\mc{S}\subseteq\mc{N}}\;\; \mc{O}(\mc{S}),
	\;\;\; \subjectto\,\,\; \; \abs{\mc{S}} = r.
\end{equation}

In the context of SP in WDNs, solving $\mathbf{P1}$ refers to finding the best sensor configuration $\mc{S}^*$ containing $r$ sensors to be allocated at the network nodes $n_L$, while an observability-based measure, defined by the set function $\mc{O}(\mc{S})$, is maximized. That being said, for the development of our approach, we obtain a closed-form expression for~\eqref{eq:obs_gramian_x0} under the sensor parameterization implied in $\mc{S}$. The following expression represents the parameterized observability Gramian~\eqref{eq:obs_gramian_x0} for multi-species WQ models~\eqref{eq:DT_dynamics_compact} with parameterized measurement model~\eqref{eq:DT_measurement_model} around a particular initial state $\hat{\vx}_0$ and with $\mc{S}\subseteq\mc{N}$ can be written as 
\begin{subequations}\label{eq:param_obs_gramian}
	\begin{align}
		\m W(\mc{S},\hat{\vx}_0) &= \sum_{j=1}^{n_y} \gamma_j 	\left(\sum_{i=0}^{N_s-1}\left(\dfrac{\partial \hat{\vx}_i}{\partial \hat{\vx}_0}\right)^{\hspace{-0.1cm}\top} \hspace{-0.075cm}\m c_j^\top \m c_j \dfrac{\partial \hat{\vx}_i}{\partial \hat{\vx}_0}\right),\\
		&= \sum_{j\in \mc{S}} \left(\sum_{i=0}^{N_s-1}
		\left(\dfrac{\partial \hat{\vx}_i}{\partial \hat{\vx}_0}\right)^{\hspace{-0.1cm}\top} 	\hspace{-0.075cm}\m c_j^\top \m c_j \dfrac{\partial \hat{\vx}_i}{\partial \hat{\vx}_0}\right),
	\end{align}
\end{subequations}	
where $\m c_j\in\mbb{R}^{1\times n_x}$ is the $j$-th row of $\m C$. Readers are referred to Appendix~\ref{apndx:proofs} for the derivation of observability Gramian~\eqref{eq:param_obs_gramian}. We note that the notation $j\in \mc{S}$ corresponds to every activated sensor such that $\gamma_j = 1$.

In WDNs, flow conditions primarily depend on the consumer’s water demand pattern, along with other static variables such as connecting components and pipe characteristics (i.e., diameter, length, material). Consequently, such flows affect the hydraulic conditions and thus the chlorine and other disinfectant byproducts within the network. To observe such changes in WQ concentrations within the network under varying hydraulic conditions, sensors should be optimally allocated to capture the underlying reactivity modeling between WQ species dynamics that are affected by such hydraulic conditions. Notice that the introduced observability Gramian~\eqref{eq:param_obs_gramian} takes into account the variations of WQ dynamical states within the network. The multi-species interactions and their dependency on hydraulic conditions are incorporated into the SP optimization problem through $\tfrac{\partial \hat{\vx}_i}{\partial \hat{\vx}_0}$. This state-space vector depicts the variation of the WQ concentrations within the network for the duration $\Delta t_{H}$ of the hydraulic simulations. 

\subsection{Robustification of the SP Observability Measures}\label{sec:derivation_SP_MSWQ_measures}
To account for the impact of changing hydraulic conditions on the optimal sensor configuration $\mc{S}^{}$ resulting from solving $\mathbf{P1}$, we define $H_{d_i} \in \mathbb{R}^{n_{\mr{N} \times N_s-1}}$ for all $i \in n_{\mr{N}}$ network nodes during a distinct $\Delta t_{H}$ hydraulic simulation. It is important for the SP problem $\mathbf{P1}$ to account not only for the varying daily demand patterns but also to yield a single optimal sensor configuration $\mc{S}^{}$ for common demand patterns $H_{d_i} \in \mc{H}_{d}$ identified by system operators. The demand patterns $H_{d_i}$ affect the hydraulic profile and, consequently, the state-space vector $\tfrac{\partial \hat{\vx}_i}{\partial \hat{\vx}_0}$. To that end, let $\kappa \in \{1,2,\cdots,d\}$ define the index of $d$ different common demand patterns $H_{d_i} \in \mc{H}_{d}$ identified by WDN operators and let $\hat{\vx}_0^{(\kappa)}$ define the system WQ concentrations resulting from the presumed hydraulic conditions identified as $\mc{H}_d$. Then, we introduce the following observability-based objective function $\mc{O}\left(\mc{S}\right)= \mc{O}\left(\mc{S}, \mc{H}_d \right)$ that accounts for the varying demand patterns that can exist in the daily operation of WDNs
\begin{equation}\label{eq:objective_fxn}
	\mc{O}\left(\mc{S}, \mc{H}_d \right):=\frac{1}{d}\sum_{\kappa=1}^{d}\mc{L}\left({\m W}^{(\kappa)}(\mc{S},\hat{\vx}_0^{(\kappa)}) \right),
\end{equation}
where $\mc{L}(\cdot)$ is an appropriate observability measure function mapping a matrix to a scalar, and the observability Gramian representing the different demand patterns is defined as
\begin{equation}\label{eq:param_obs_gramian_kappa}
		\hspace{-0.15cm}{\m W}^{(\kappa)}(\mc{S},\hat{\vx}_0^{(\kappa)}) := 
		\sum_{j\in \mc{S}} \left(\sum_{i=0}^{N_s-1}
		\left(\dfrac{\partial \hat{\vx}_i^{(\kappa)}}{\partial \hat{\vx}_0^{(\kappa)}}\right)^{\hspace{-0.1cm}\top} 	\hspace{-0.075cm}\m c_j^\top \m c_j \dfrac{\partial \hat{\vx}_i^{(\kappa)}}{\partial \hat{\vx}_0^{(\kappa)}}\right).
\end{equation}	

Note that we changed the notation of $\mc{O}\left(\mc{S}, \mc{H}_d \right)$ to account for the different hydraulic and demand profiles considered. As such, the notation $\mc{O}\left(\mc{S}, \mc{H}_d \right)$ is equivalent to $\mc{O}\left(\mc{S}\right)$ but provides more clarity. Furthermore, the validity of the submodular set properties for the above observability-based measure is supported by the submodular-preserving operation, conic combination~\cite{Bach2013b}. That is, original submodular function properties are retained under a non-negative weighted sum. To that end, we consider two submodular scalar mapping functions $\mc{L}(\cdot)$: $\mr{trace}$ and $\mr{log\,det}$, for quantifying the observability of WQ dynamics within WDNs. When the function $\mc{L}(\cdot)$ for objective function~\eqref{eq:objective_fxn} used in the SP problem $\mathbf{P1}$ takes the form of the $\mr{trace}$ and $\mr{log\,det}$, the robust measures that account for varying demand profiles are said to be modular and submodular. Specifically, the $\mr{trace}$ of the observability Gramian~\eqref{eq:param_obs_gramian_kappa} under such a conic combination is said to be modular, as such the SP objective function $\mc{O}\left(\mc{S}, \mc{H}_d \right)$ considering the $\mr{trace}$ can be written as follows
\begin{align}\label{eq:trace_mod} 
	\mc{O}_{\mr{trace}}:=\frac{1}{d}\sum_{\kappa=1}^{d}\mr{trace}\left({\m W}^{(\kappa)}(\mc{S})\right),
\end{align}
while the $\mr{log\,det}$ for the conic combination for $\kappa \in \{1,2,\cdots,d\}$ is said to be submodular and monotone increasing, thus the objective function $\mc{O}\left(\mc{S}, \mc{H}_d \right)$ can be written as follows
\begin{align}\label{eq:logdet_submodular} 
		\mc{O}_{\mr{log\,det}}:=\frac{1}{d}\sum_{\kappa=1}^{d}\mr{log\,det}\left({\m W}^{(\kappa)}(\mc{S})\right).
\end{align} 

The derivation and validity of the observability measures for multi-species WQ dynamics under varying demand profiles are shown in Appendix~\ref{apndx:proofs}. 

\subsection{Robust SP Algorithm in Nonlinear WQ Systems}\label{sec:SP_MSWQ}
For small WDNs, the optimal sensor configuration $\mc{S}^*$, obtained from solving $\mathbf{P1}$ under $\mc{O}_{\mr{trace}}$ and $\mc{O}_{\mr{log\,det}}$, can be determined using a brute-force approach. However, for larger-scale networks, this becomes computationally infeasible. This type of optimization problem is known to have large computational complexity, while there is no polynomial-time algorithm that guarantees reaching the optimal solution. To address this issue, we employ the fact that the presented observability measures render $\mathbf{P1}$ submodular and monotone increasing, allowing us to apply the greedy algorithm to efficiently determine sensor configuration.

The computational framework for solving $\mathbf{P1}$ using greedy algorithms results in desirable performance while being computationally feasible. That being said, for a submodular and monotone increasing set function $\mc{O}(\cdot)$, the greedy algorithm offers the following performance guarantee~\cite{Nemhauser1978}.
\begin{equation}\label{eq:performance-SNS}
	\mc{O}^*_{\mc{S}} -\mc{O}(\emptyset) \geq \left(1-\tfrac{1}{e}\right)\left(\mc{O}^*-\mc{O}(\emptyset)\right), 
	\;\; \text{with}\; \mc{O}(\emptyset) =0,
\end{equation}
where $\mc{O}^*_{\mc{S}}$ is the optimal value of $\mathbf{P1}$ and $e\approx 2.71828$. This means that the solution to the WQ sensor placement problem is at least $63\%$ optimal. Note that a plethora of literature~\cite{Krause2011, Summers2016,Tzoumas2016, Zhou2019,Bilmes2022} has shown that the solution can achieve up to $99\%$ accuracy, and that this guarantee is merely a theoretical bound. We can now apply a greedy algorithm~\cite[Algorithm 1]{Kazma2023f} to solve the multi-species WQ sensor placement problem while accounting for various demand profiles in WDNs. The detailed framework for computing the optimal SP configuration is provided in Algorithm~\ref{algorithm1}.

\begin{algorithm}[h]
	\caption{Robust optimal SP configuration for nonlinear multi-species WQ models of WDNs.}\label{algorithm1}
	\DontPrintSemicolon
	\textbf{input:} identified demand profiles $H_{d_i} \in \mc{H}_{d}$, water network parameters (WQ and hydraulic parameters), objective function~\eqref{eq:objective_fxn}\;
	\textbf{initialize:} sensor configuration ${\mc{S}}=\emptyset$, $k\leftarrow0$, $j\leftarrow1$ \hspace{-0.5cm}\tcp*{initialize time-step and sensor counters}
	\textbf{compute:} observation horizon $N_s= \tfrac{T_s}{\Delta t_{\mr{WQ}}}$\;
	\tcc{Loop over demand and different WQ condition scenarios}
	\For{$\kappa \in \{1,2,\cdots,d\} $} {
		// \textit{for each demand profile simulation $\kappa$}\;
		\For{$k \leq N_s$}{
			\textbf{simulate:} WQ dynamics~\eqref{eq:DT_dynamics_compact}\;
			\textbf{compute:} ${\m W}^{(\kappa)}(\mc{S},\hat{\vx}_0^{(\kappa)})$~\eqref{eq:param_obs_gramian_kappa}}
	}
	\tcc{Greedy sensor selection loop}
	\While{$j \leq r$}{ 
		\textbf{choose:} $\mc{O}\left(\mc{S}, \mc{H}_d \right) = \mc{O}_{\mr{trace}} \text{ or } \mc{O}_{\mr{log\,det}}$ \;
		\tcp{marginal grain from each candidate sensor}
		\textbf{compute:} $\mc{G}_j = \mc{O}(\mc{S}_j\cup \{a\})-\mc{O}(\mc{S}_j)$, $\forall a\in \mc{N}\setminus \mc{S}_j$ \;
		\textbf{assign:} $\mc{S}_j\leftarrow \mc{S}_j \cup \left\{\mathrm{arg\,max}_{a\in \mc{N}\setminus \mc{S}_j}\,\mc{G}_j \right\}$ \tcp*{add max-gain sensor node}
		\textbf{update:} $j \leftarrow j + 1$\;
	}
	\textbf{output:} optimal sensor configuration computed as
	$\mc{S}^{*} \leftarrow {\arg\max}_{\mc{S} \subseteq \mc{N}, \,\, |\mc{S}| = r} \;\; \mc{O}(\mc{S})$ \; 
\end{algorithm}

The inputs for the algorithm are the choice of sensor number $r \leq n_{\mr{N}}$, identified demand profiles $H_{d_i} \in \mc{H}_{d}$, and both hydraulic and WQ parameters. The observation horizon $N_s$ is computed based on the total simulation time and the WQ time-step $\Delta t_{\mr{WQ}}$. The algorithm is initialized by setting the sensor configuration set $\mc{S}$ to the empty set $\emptyset$. The notation $\mc{S}_j$ denotes the sensor configuration set with $j$ selected sensors at iteration $j \leq r$. Variable $j \in \mc{S}$ defines an element (i.e., a node) in the set $\mc{S}$. For each demand scenario $\kappa$, the nonlinear WQ dynamics are simulated, and an observability Gramian is constructed at each time-step. The algorithm greedily and incrementally updates the sensor set $\mc{S}_j$ until the sensor budget constraint, $|\mc{S}| = r$, is met. The output of Algorithm~\ref{algorithm1} is the robust SP configuration $\mc{S}^*$. The observability-based measures for solving can be chosen according to~\eqref{eq:trace_mod} and~\eqref{eq:logdet_submodular}. The algorithm terminates by computing the final optimal sensor configuration $\mc{S}^*$ by choosing sensors that achieve maximum marginal gain with respect to the chosen observability measure that ensure optimal state estimation performance as discussed in Section~\ref{sec:Obs-WQD}. That is, the algorithm chooses a single sensor configuration $\mc{S}^*$ that allows observing and estimating the WQ dynamics (depicting chlorine and other reactants) throughout the WDN while considering varying water demand profiles.

\section{Case Studies}\label{sec:casestudy}
In this section, we illustrate and validate the proposed multi-species WQ sensor placement framework on two WDNs. The objectives of the case studies are outlined as follows.
\begin{itemize}[leftmargin=*]
	\item To illustrate that the proposed SP Framework accounts for varying water demand patterns and WQ conditions/parameters (Sections~\ref{sec:impact_hydro} and~\ref{sec:impact_WQ}). This is essential in order to show evidence that the framework depicts the nonlinear WQ dynamics.
	\item To provide a single robust sensor configuration to account for the aforementioned variability and nonlinearities in WQ modeling (Sections~\ref{sec:main_result} and~\ref{sec:main_result_NET2}). This enables system operator to allocate a single sensor configuration that accounts such variability in network operation.
	\item To showcase the scalability of the proposed SP framework under different network parameters and scales (Sections~\ref{sec:main_result_NET2} and~\ref{sec:compt_time}). This provides computational evidence regarding the applicability of the proposed method to larger scale networks.
\end{itemize}

\begin{figure}[t]
	\centering
	\subfloat{\includegraphics[keepaspectratio=true,scale=0.7]{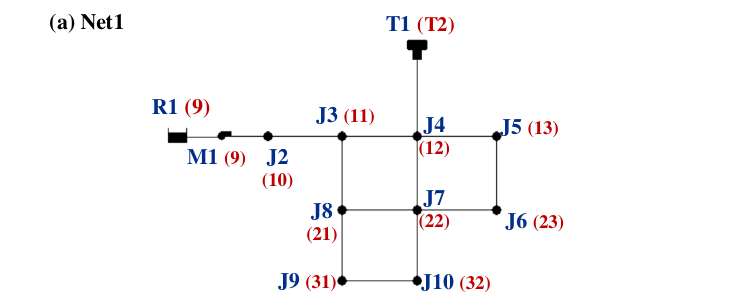}}{}{}\vspace*{-0.45cm}
	\subfloat{\includegraphics[keepaspectratio=true,scale=0.875]{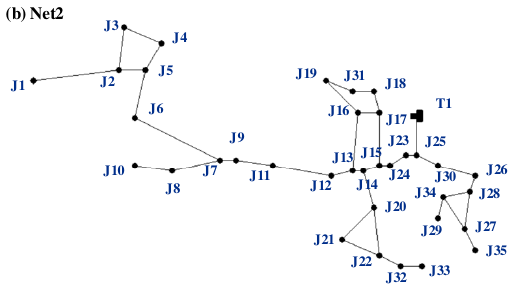}}
	\caption{Network typologies and component labels for the considered networks: (a) Net1 and (b) Net2. The {\color[rgb]{0.753,0,0} red color} indicates labels for Net1 in~\cite{Taha2021c}.}\label{fig:networks2}
	\vspace{-0.6cm}
\end{figure}

The networks considered, Net1 and Net2, are depicted in Fig.~\ref{fig:networks2}, and the components of each network are summarized in Table~\ref{tab:WDN_components}. In Net2, junction J1 represents a pump station and accordingly, is simulated with a negative demand indicating the water supply of the station into the system. The two networks are benchmark networks and are considered representative of real-world WDNs, with topologies that include loops and dead-end branches. These networks are widely used in the literature for WQ modeling and SP applications; see~\cite{Vrachimis2019, Taha2021c, Seyoum2017, Elsherif2023a}. Note that the networks chosen are of contrasting size and topology, with Net2 being larger and exhibiting more sub-networks and dead-end branches compared to the looped structure of Net1. The networks are studied under different hydraulic settings and scenarios, and initial WQ conditions. The hydraulic settings are obtained using the EPANET toolkit on MATLAB. The WDN simulation and solution of the submodular maximization problem $\mathbf{P1}$ are performed in MATLAB R2024b, running on a MacBook Pro with an Apple M1 Pro chip, a 10-core CPU, and 16 GB RAM.

\begin{table}[b]
		\vspace{-0.2cm}
	\centering 
	\caption{Considered WDNs: summary of components}
	\label{tab:WDN_components}
	\vspace{-0.2cm}
	\renewcommand{\arraystretch}{1.3}
	\begin{tabular}{l|c|c|c|c|c}
		\midrule \hline
		\textbf{Network} & Junctions & Reservoirs & Tanks & Pipes & Pumps \\ \hline
		\textbf{Net1}    & $9$         & $1$          & $1$     & $12$    & $1$     \\ \hline
		\textbf{Net2}   & $35$        & $0$          & $1$     & $40$   & $0$     \\ \hline
		\toprule \bottomrule
	\end{tabular}
\end{table}

\begin{figure}[t]
	\subfloat{\includegraphics[keepaspectratio=true,scale=0.4]{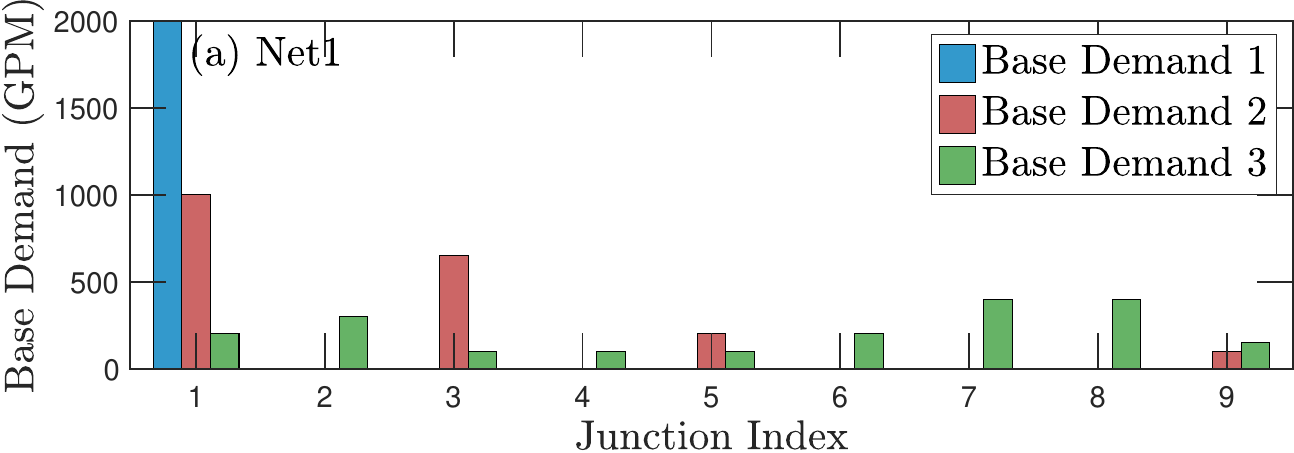}}{}{}	\hspace*{+0.15cm}
	\subfloat{\includegraphics[keepaspectratio=true,scale=0.4]{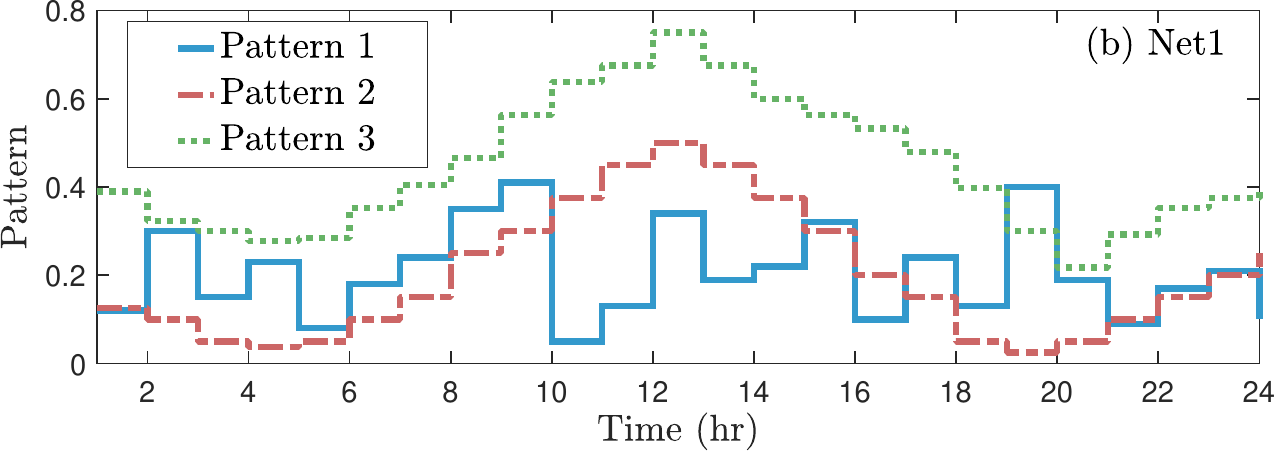}}
	\caption{(a) Junction demand for each of the three cases considered and (b) the respective hydraulic pattern for each case in Net1.}\label{fig:demand}
	\vspace{-0.5cm}
\end{figure}

For Net1, we consider three different hydraulic scenarios, for which Fig.~\ref{fig:demand} illustrates the base demands at each junction of the network and the corresponding demand pattern. We refer to these scenarios as cases 1, 2, and 3. For all these cases, the WQ time-step is taken to be $\Delta t_\mathrm{WQ} = 10$ secs, while the hydraulic time-step is $\Delta t_\mathrm{H} = 1$ hr. In addition, the initial WQ conditions $\vx_{0}\in\mc{X}_0$ are 2 mg/L and 0.3 mg/L for chlorine and the fictitious reactant at R2, and zero elsewhere. Note that in this work, we model demand using a piecewise-constant pattern, where consumers' demand is assumed to be constant within each hydraulic time-step (one hour), and any variation is captured at the transitions between steps. While in real-world water networks demand changes continuously in real time, the adopted piecewise-constant representation aligns is considered sufficient for capturing variability at the time-scale relevant to WQ modeling and sensor allocation applications; see~\cite{Seyoum2017, Taha2021c, Elsherif2023a}. The continuously changing demand patterns can be accounted for by reducing the hydraulic time-step to match with the resolution of any existing demand pattern.

\begin{figure}[b]
	\vspace{-0.5cm}
	\subfloat{\includegraphics[keepaspectratio=true,scale=0.4]{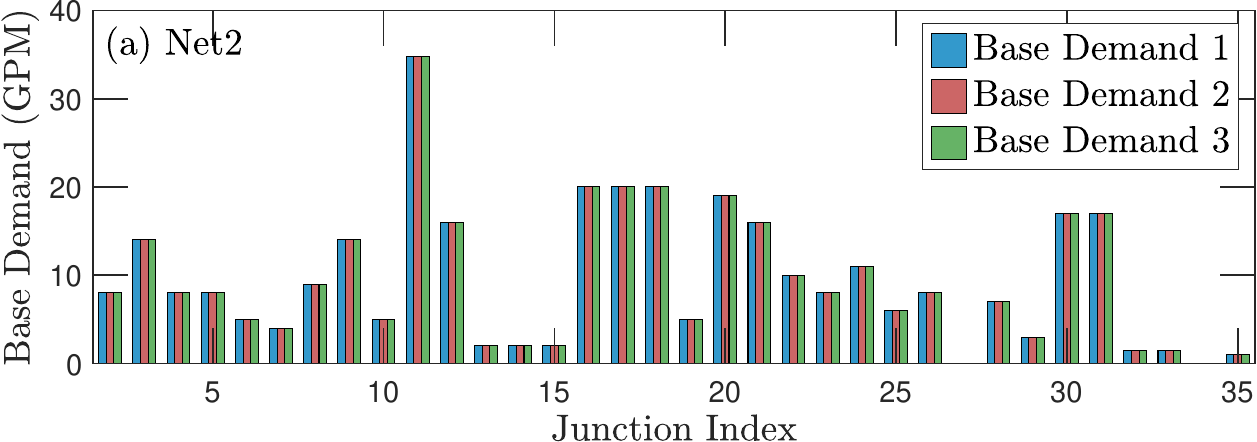}~\label{fig:demand2a}}{}{}
	\subfloat{\includegraphics[keepaspectratio=true,scale=0.4]{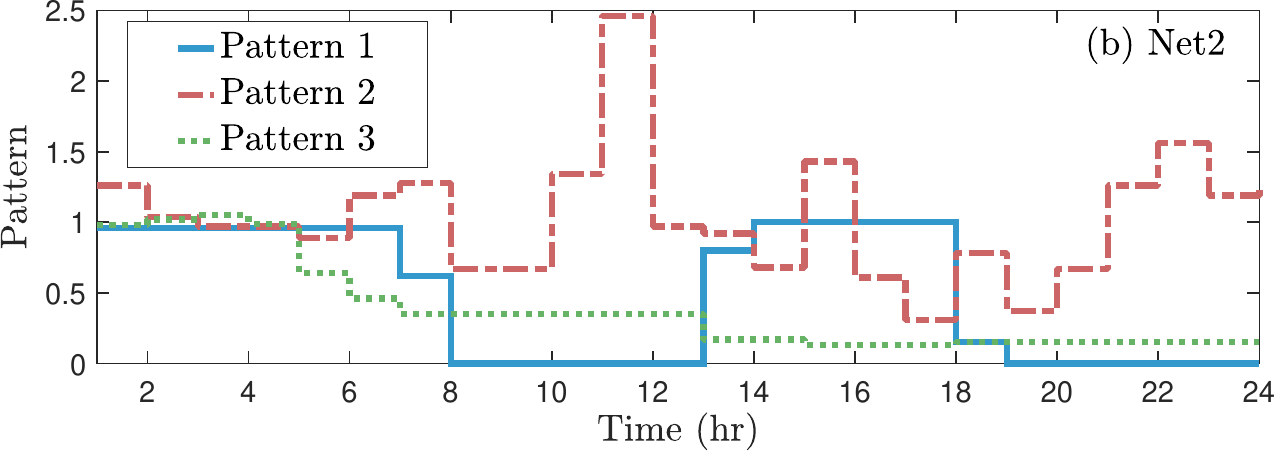}}
	\caption{(a) Junction demand for each of the three cases considered and (b) the respective hydraulic patterns for each case in Net2.}\label{fig:demand2}
\end{figure}

For Net2, three different scenarios are included in the sensor placement problem. In these scenarios, junctions have the same base demand but different patterns, as demonstrated in Fig.~\ref{fig:demand2}. Note that the demand at J1 is excluded from Fig.~\ref{fig:demand2a} as it presents a pseudo-negative demand, as mentioned earlier. The WQ time-step for these scenarios is $\Delta t_\mathrm{WQ} = 30$ secs within a hydraulic time-step of $\Delta t_\mathrm{H} = 1$ hr. The WQ initial conditions for the first scenario are 1.5 mg/L of chlorine and 0.3 mg/L of fictitious reactant at J1, and 1 mg/L of chlorine and 0.1 mg/L of fictitious reactant at TK1, with zeros elsewhere. For the second scenario, the chlorine concentration at J1 is updated to 2 mg/L, while all initial conditions are kept the same for the third scenario. The mutual reaction coefficient for the first case is set equal to the bulk decay coefficient; it is doubled for the second scenario and tripled for the third. Note that for both networks, the WQ time-step is chosen to ensure numerical stability of the discretization scheme used to simulate WQ dynamics. Typically, in standard EPANET-MSX-based simulations~\cite{Seyoum2017}, this value can be as large as 5 minutes. The hydraulic time-step is slower and typically set to 1 hour with a 24-hour observation horizon, therefore representing a full daily cycle of water demand. The time-step choices thus represent a realistic basis for the SP framework.

 \begin{figure*}[t]
	\subfloat{\includegraphics[keepaspectratio=true,scale=0.52]{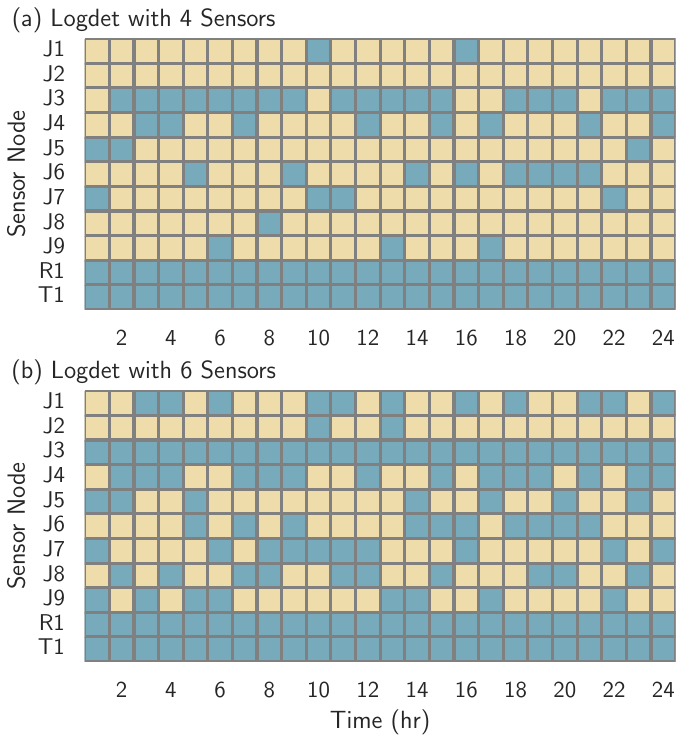}}
	\subfloat{\includegraphics[keepaspectratio=true,scale=0.52]{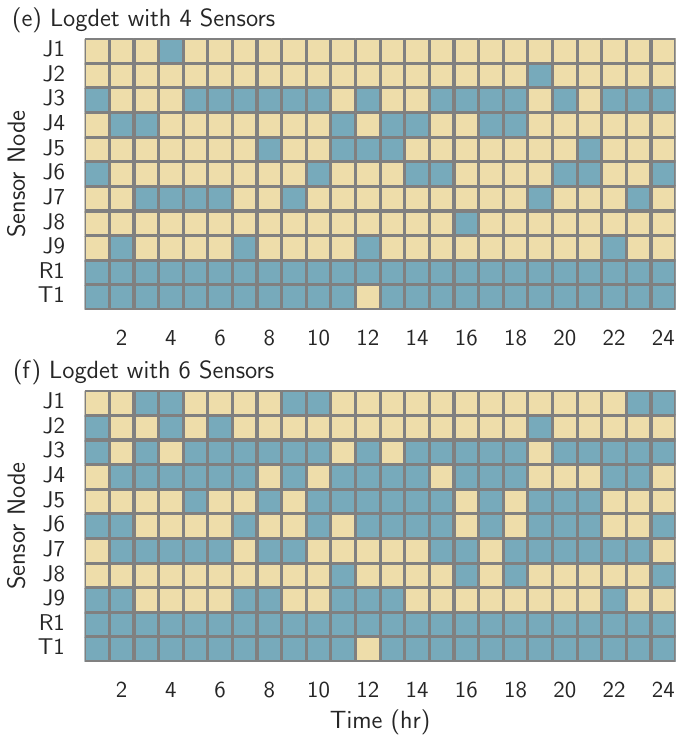}}
	\subfloat{\includegraphics[keepaspectratio=true,scale=0.52]{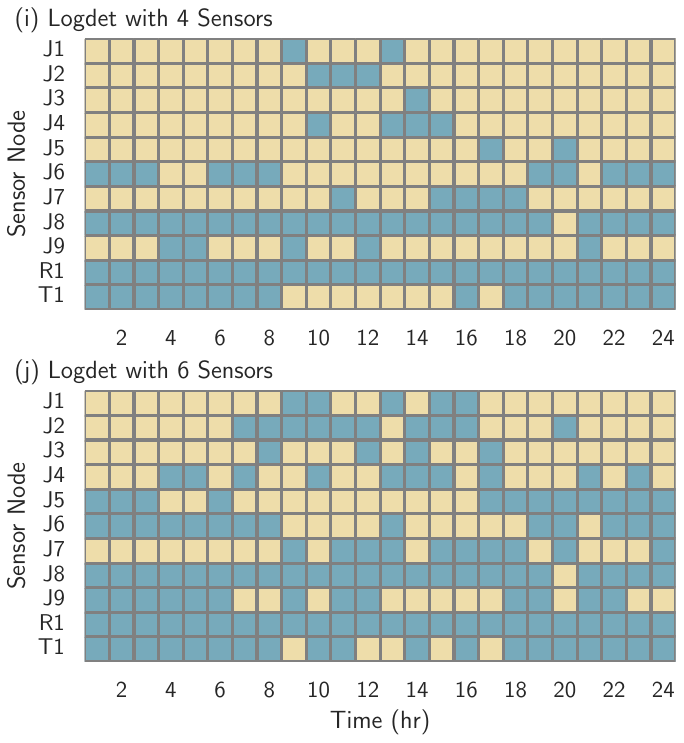}}{}{}
	\subfloat{\includegraphics[keepaspectratio=true,scale=0.52]{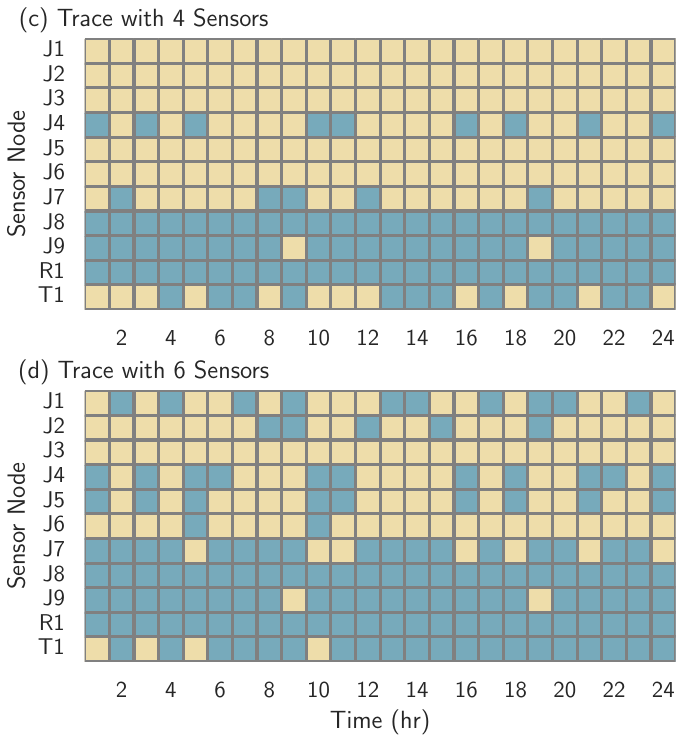}}
	\subfloat{\includegraphics[keepaspectratio=true,scale=0.52]{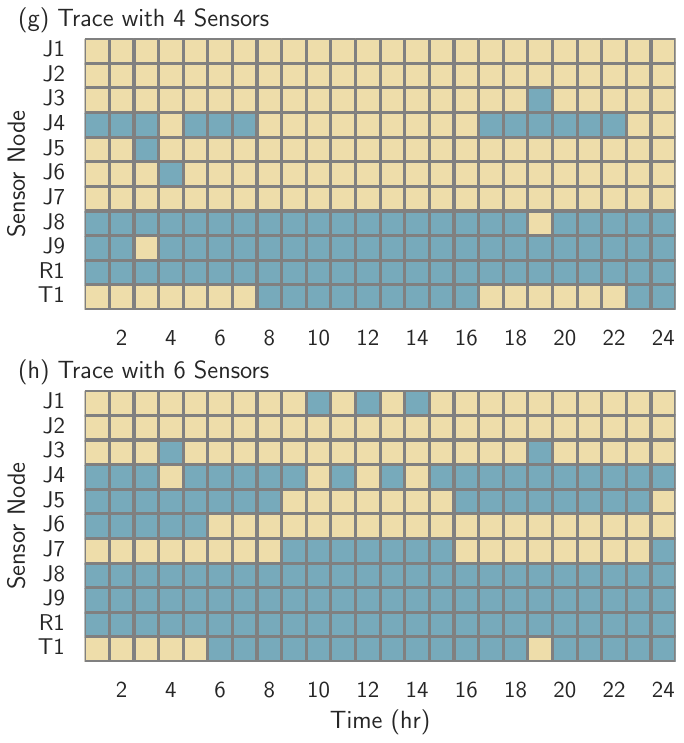}}
	\subfloat{\includegraphics[keepaspectratio=true,scale=0.52]{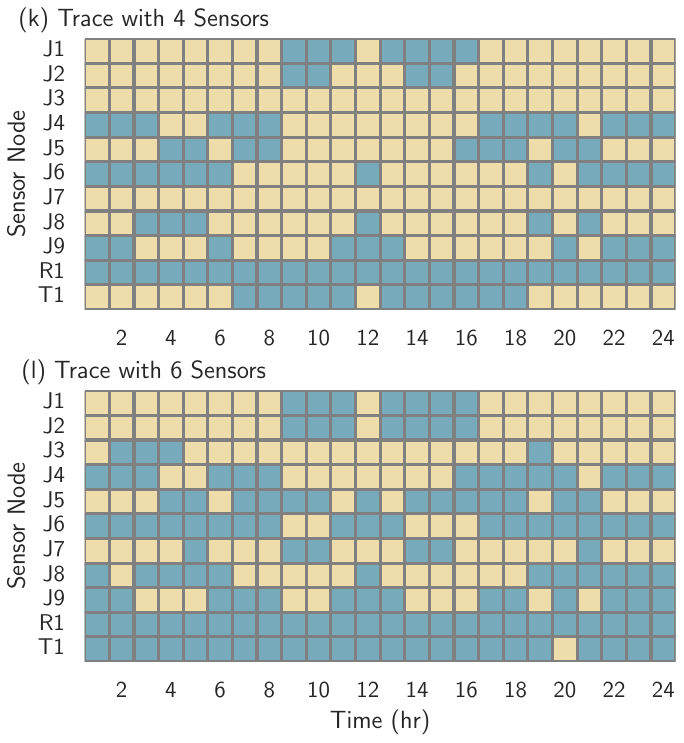}}{}{}
	\centering 
	\caption{Optimal SP results under varying water demand patterns and hydraulic profiles for Net1. The figure shows configurations using two objective functions: $\mc{O}_{\mr{log\,det}}$ and $\mc{O}_{\mr{trace}}$, with sensor configurations of cardinality 4 and 6. Figures (a,b,e,f,i,j) depict configurations for $\mc{O}_{\mr{log\,det}}$, while figures (c,d,g,h,k,l) depict configurations for $\mc{O}_{\mr{trace}}$. Each row represents a specific sensor configuration, with 4 sensors on the first row and 6 sensors on the second. Columns 1, 2, and 3 correspond to cases 1, 2, and 3, respectively. Each figure shows the SP for each hydraulic time-step $\Delta t_H$ across the WDN potential sensor nodes. The {\color[rgb]{0.93,0.87,0.67} light color} indicates nodes without sensors, while the {\color[rgb]{0.47,0.67,0.73} dark color} represents sensor nodes.}\label{fig:NET1-results}
	\vspace{-0.4cm}
\end{figure*}

\subsection{Impact of Demand and Hydraulic Patterns}\label{sec:impact_hydro}
Having presented the above case studies, we now outline the main findings of the proposed SP optimization framework. Fig.~\ref{fig:NET1-results} showcases the optimal SP results obtained by solving $\mathbf{P1}$ for each of the aforementioned scenarios individually, i.e., for $\kappa=1$. The results show that the optimal sensor configuration $\mc{S}^{*}$ varies under different hydraulic scenarios and along the hourly hydraulic patterns. Furthermore, the results also show that utilizing different objective functions for $\mathbf{P1}$, defined as $\mc{O}_{\mr{log\,det}}$~\eqref{eq:logdet_submodular} and $\mc{O}_{\mr{trace}}$~\eqref{eq:trace_mod}, yields distinct sensor configurations $\mc{S}^{*}$. For Net1, the SP problem is solved for two sensor configuration of cardinality $|\mc{S}_1| =4$ and $|\mc{S}_2| =6$. We note here that by comparing the sensor configurations for each of the hydraulic time-steps under the two sensor numbers considered, the sensor configuration $\mc{S}_2$ is given by $\mc{S}_2 = \mc{S}_1 \bigcup \{r_5,\; r_6\}$. This means that the set of sensors obtained when solving for a configuration of $4$ sensors is included in that obtained for $6$ sensors, with only two additional sensors, ${r_5, r_6}$, added to the previous solution. These results validate the modular and submodular properties of the observability measures that define the objective function for $\mathbf{P1}$. The importance of this result can be highlighted when WDN operators want to consider allocating additional sensors in a preexisting WDN. By considering the proposed control-theoretic approach the system operators can solve for the additional sensor required without the need to remove or re-allocated prior existing WQ sensors.

\begin{figure*}[t]
	\subfloat{\includegraphics[keepaspectratio=true,scale=0.52]{LogDet_CS1.pdf}}
	\subfloat{\includegraphics[keepaspectratio=true,scale=0.52]{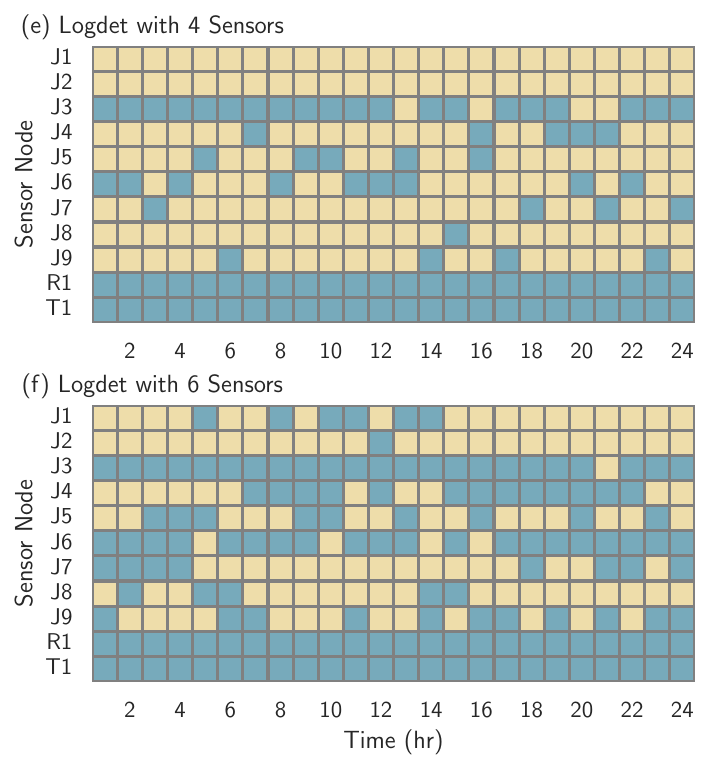}}
	\subfloat{\includegraphics[keepaspectratio=true,scale=0.52]{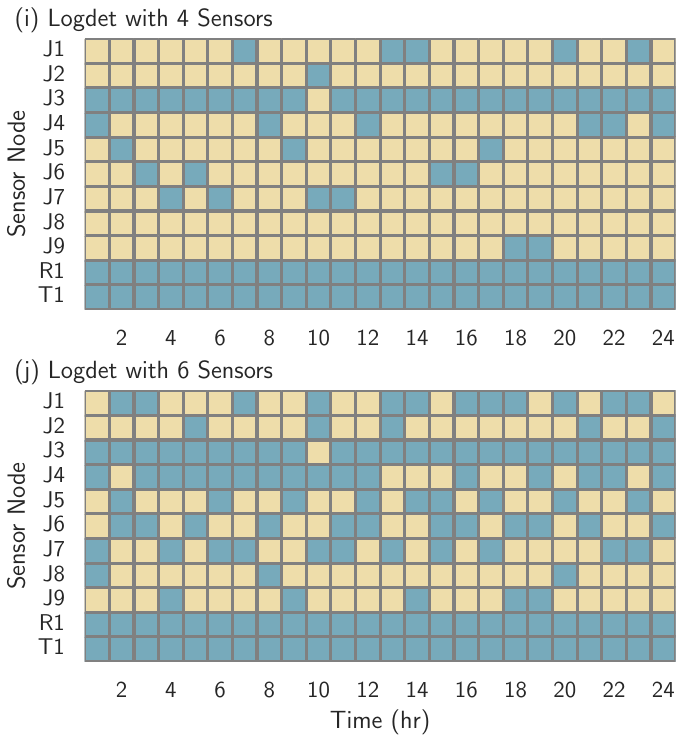}}{}{}
	\subfloat{\includegraphics[keepaspectratio=true,scale=0.52]{Trace_CS1.pdf}}
	\subfloat{\includegraphics[keepaspectratio=true,scale=0.52]{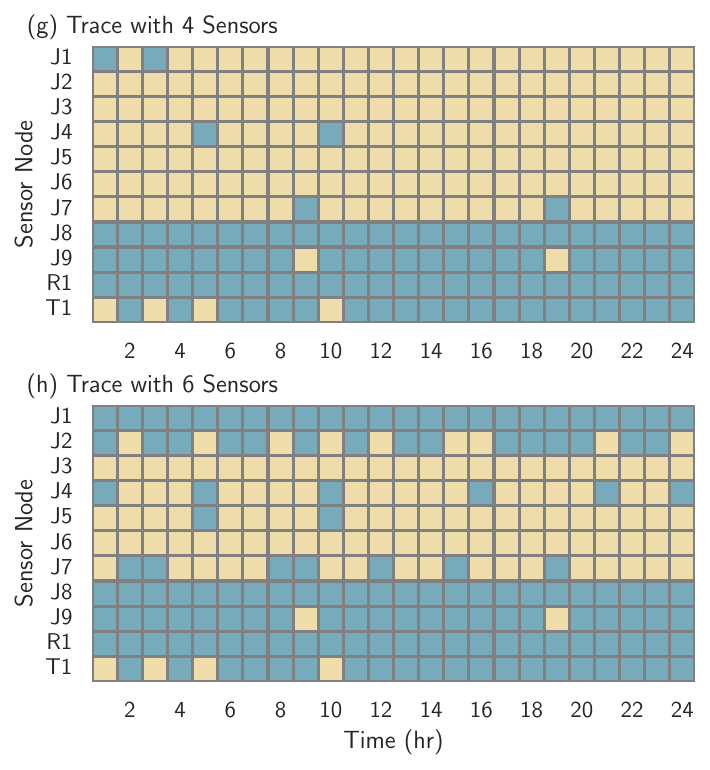}}
	\subfloat{\includegraphics[keepaspectratio=true,scale=0.52]{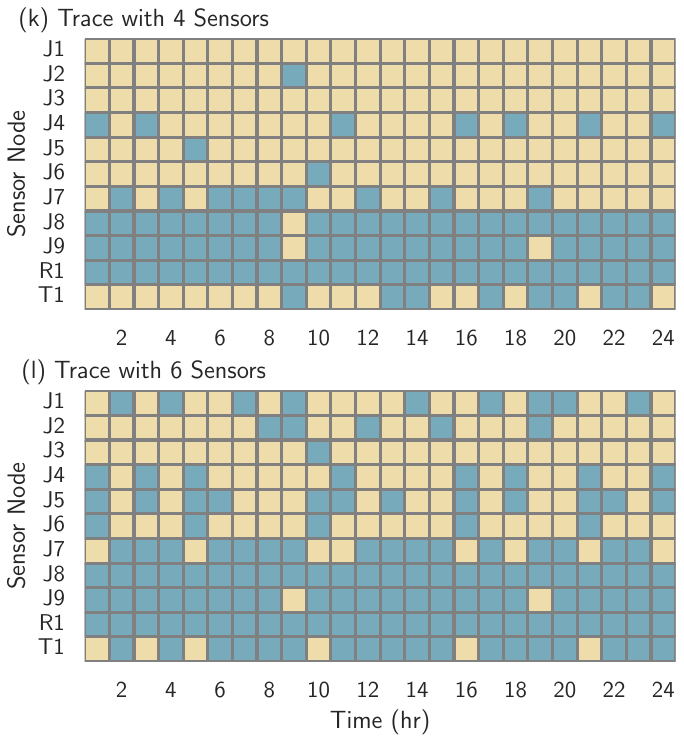}}{}{}
	\centering 
	\caption{Optimal SP results under varying WQ conditions for Net1. The figure shows configurations using two objective functions: $\mc{O}_{\mr{log\,det}}$ and $\mc{O}_{\mr{trace}}$, with sensor configurations of cardinality 4 and 6. Figures (a,b,e,f,i,j) depict configurations for $\mc{O}_{\mr{log\,det}}$, while figures (c,d,g,h,k,l) depict configurations for $\mc{O}_{\mr{trace}}$. Each row represents a specific sensor configuration, with 4 sensors on the first row and 6 sensors on the second. Columns 1, 2, and 3 correspond to cases 1, 4, and 5, respectively. Each figure shows the SP for each hydraulic time-step $\Delta t_H$ across the WDN potential sensor nodes.}\label{fig:NET1-resultsMSX}
\end{figure*}

\begin{figure}[t]
	\centering
	\includegraphics[keepaspectratio=true,scale=0.51]{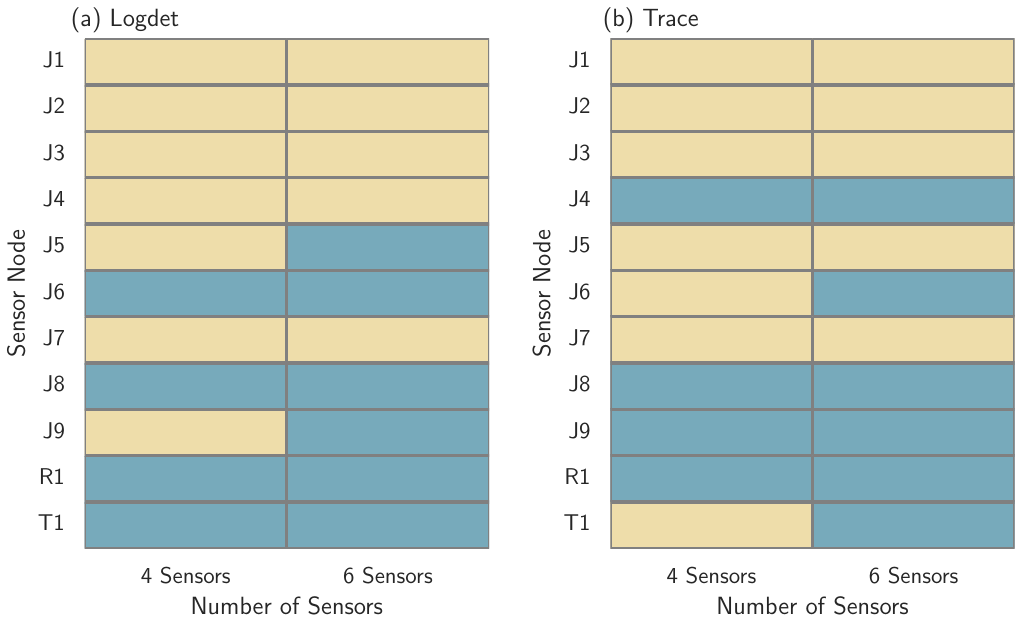}
	\caption{Optimal sensor configuration $\mc{S}^{*}$ obtained by solving $\mathbf{P1}$ using Algorithm~\ref{algorithm1} with $\kappa=5$ for Net1. The result in (a) is for the objective function $\mc{O}_{\mr{log\,det}}$, and in (b) for $\mc{O}_{\mr{trace}}$.}\label{fig:NET1-OSP}
		\vspace{-0.3cm}
\end{figure}

\subsection{Impact of Varying WQ Conditions/Parameters}\label{sec:impact_WQ}
In addition to the previous cases for Net1, we consider two additional scenarios, referred to as cases 4 and 5. In these scenarios, we modify the multi-species dynamics in the system to observe how these changes affect sensor allocations. In case 4, the mutual reaction coefficient $\alpha_r$ is reduced to half its value in case 1, and doubled for case 5. Moreover, the initial condition for the fictitious reactant is changed for both scenarios to be 0.35 mg/L at R2 and 0.1 mg/L elsewhere. The hydraulic profiles for these two cases are the same as in case 1, as well as the hydraulic and WQ time-steps. 

The results for solving $\mathbf{P1}$ for each of the WQ dynamics cases are depicted in Fig.~\ref{fig:NET1-resultsMSX}. Similarly to varying hydraulic conditions, the optimal SP problem is able to account for varying multi-species WQ initial conditions and different decay parameters. This result extends the previous findings of a control-theoretic approach that was developed for single-species WQ dynamics in~\cite{Taha2021c}. As such, by accounting for the multi-species WQ reactivity modeling of chlorine with other reactants, the SP problem can be extended to address contaminant intrusion events and more typical conditions resulting from WQ parameters such as pipe aging and biofilm development. We note that in Fig.~\ref{fig:NET1-resultsMSX}, the results for case 1 are also included redundantly for visualization purposes and as a means to compare among the three different cases with varying WQ dynamics (cases 1, 4, and 5). The columns represent the different cases while the first two rows are for $\mc{O}_{\mr{log\,det}}$ and the second two rows are for $\mc{O}_{\mr{trace}}$ objective functions. We note that the validity of the proposed SP framework in terms of estimating all system states for nonlinear models is validated in~\cite{Kazma2023f}.

\begin{figure*}[t]
	\centering
	\subfloat{\includegraphics[keepaspectratio=true,scale=0.54]{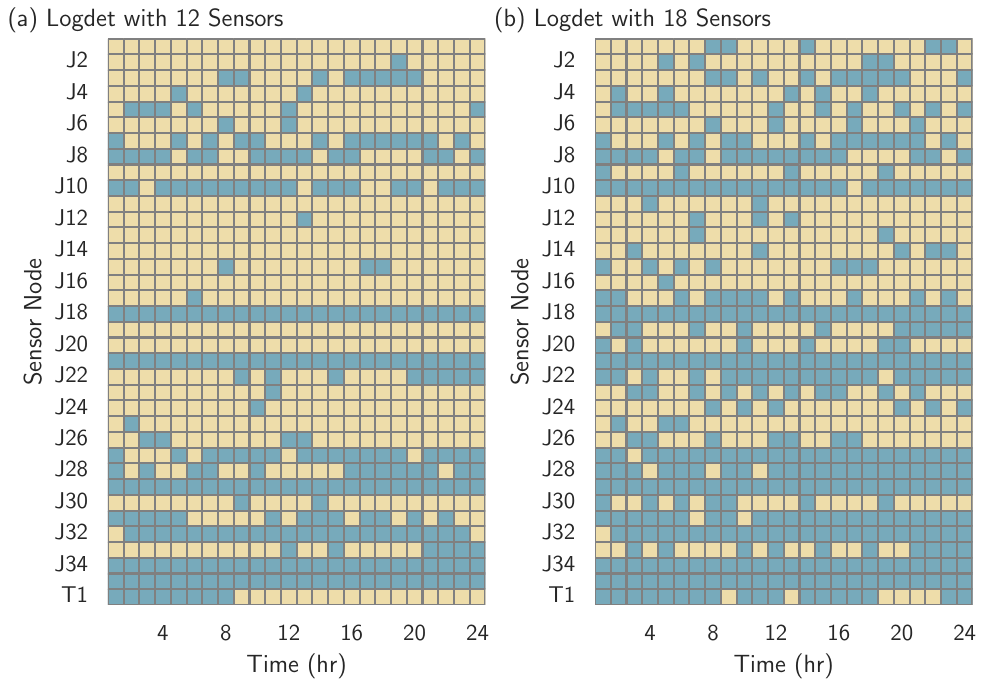}}
	\subfloat{\includegraphics[keepaspectratio=true,scale=0.54]{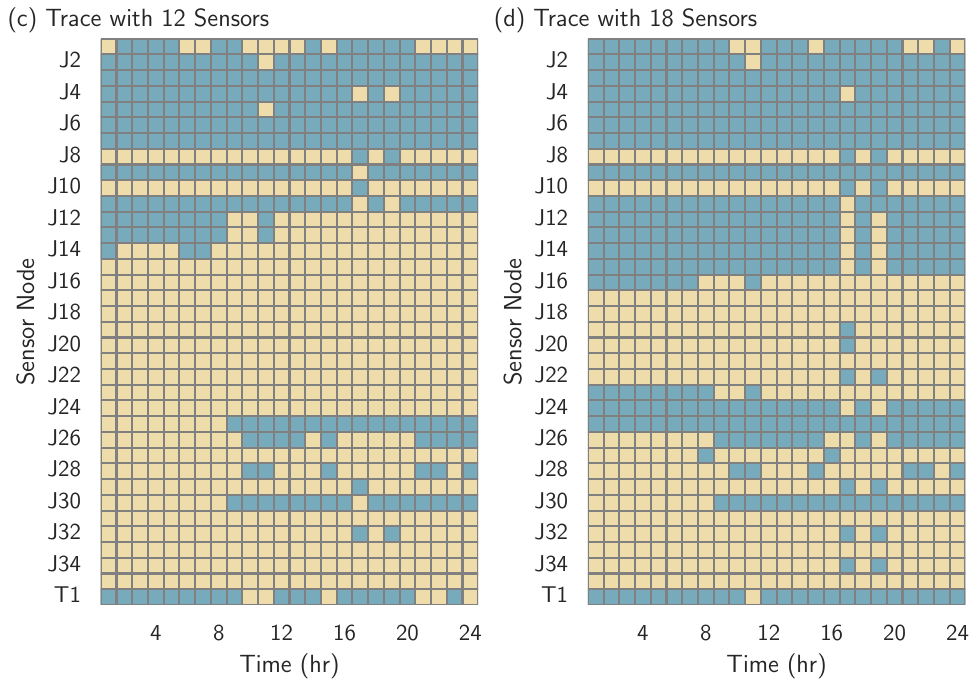}}
	\caption{Optimal SP results under varying hydraulic profile case 2 for Net2. The figure shows configurations using two objective functions: $\mc{O}_{\mr{log\,det}}$ and $\mc{O}_{\mr{trace}}$, with sensor configurations of cardinality 12 and 18. Figures (a)-(b) depict configurations for $\mc{O}_{\mr{log\,det}}$, while figures (c)-(d) depict configurations for $\mc{O}_{\mr{trace}}$. Each figure shows the optimal sensor configuration for each hydraulic time-step $\Delta t_H$ across the WDN potential sensor nodes.}\label{fig:NET2-results}
\end{figure*}

\subsection{Robust SP Result for Net1}\label{sec:main_result}
Based on the above results, we now solve $\mathbf{P1}$ by considering the 5 cases simultaneously, i.e., we solve for the optimal sensor configuration that is robust against varying hydraulic patterns and changing WQ dynamics. As such, by setting $\kappa=5$ and solving $\mathbf{P1}$ using Algorithm~\ref{algorithm1}, we obtain a singular and robust sensor configuration across the different cases and along the hourly hydraulic demand pattern. The results for the two sensor configurations, $|\mc{S}_1| = 4$ and $|\mc{S}_2| = 6$, solved for each of the two different objective functions are depicted in Fig.~\ref{fig:NET1-OSP}. Notice that, similar to the case for $\kappa=1$, by defining the observability measures under the observability Gramian~\eqref{eq:param_obs_gramian_kappa} that accounts for different cases $\kappa \in \{1,2,\cdots,d\}$, the submodular and modular properties of the objective functions $\mc{O}_{\mr{log\,det}}$ and $\mc{O}_{\mr{trace}}$ are retained. This is evident by realizing that the optimal sensor configuration set $\mc{S}^{*}_1 = \{\mr{T1}, \mr{R1}, \mr{J8}, \mr{J6}\}$ and that $\mc{S}^{*}_2 = \mc{S}^{*}_1 \bigcup \{\mr{J9},\mr{J5}\}$ when solving for $\mc{O}_{\mr{log\,det}}$. This is also true when the objective function is defined as $\mc{O}_{\mr{trace}}$, where $\mc{S}^{*}_2 = \mc{S}^{*}_1 \bigcup \{\mr{T1}, \mr{J6}\}$ and $\mc{S}^{*}_1 = \{\mr{R1},\mr{J9}, \mr{J8}, \mr{J4}\}$. 

We note that when allocating sensors within Net1, we have chosen to place a sensor on R1; therefore, R1 is always included in the optimal sensor configuration set. This showcases the ability of the system operator to impose physical constraints, such as specifying locations where sensors are always placed. This can apply to nodes or sections of the WDN that are susceptible to contamination events or are typically monitored. We note that, by observing the topology of Net1 in Fig.~\ref{fig:networks2} and the optimal SP results in Fig.~\ref{fig:NET1-OSP}, the optimal sensor configuration obtained using the $\mc{O}_{\mr{log\,det}}$ observability measure tends to allocate sensors at edge nodes, whereas the configuration from $\mc{O}_{\mr{trace}}$ allocates sensors at more central nodes. To illustrate the difference between the two observability measures and provide a perspective on their potential utilization by WDN operators, we apply the SP problem to Net2, which exhibits community structures within the network. Here, community structures are sub-networks where several nodes are connected and form a circle connecting to a nearing community, e.g., branched networks that include looped sub-networks.
\subsection{Validation of Proposed SP Framework on Net2}\label{sec:main_result_NET2}
The optimal SP results for solving $\mathbf{P1}$ for Net2 under hydraulic demand pattern 2 are illustrated in Fig.~\ref{fig:NET2-results}. For brevity, we include only the SP results under a single operational hydraulic condition and omit the rest. Notice that, similar to Net1, the optimal placement of WQ sensors depends on the hydraulic conditions. This is an important aspect that the SP framework takes into consideration, as it enables system operators to understand the impact of varying demand patterns on the validity of the sensor configuration within the network and its ability to depict WQ concentrations and detect any contamination. For this network, $\mathbf{P1}$ is solved for sensor configurations with cardinalities $|\mc{S}_1| = 12$ and $|\mc{S}_2| = 18$. The optimal SP results for each of the considered observability measures are illustrated and indeed differ. This difference highlights the distinct influence each measure has on optimal sensor configuration $\mc{S}^{*}$ result.  Meaning that, the operators choice of objective function can affect the configuration’s ability to monitor WQ dynamics and detect potential contamination within the network. 

\begin{figure}[t]
	\centering
	\includegraphics[keepaspectratio=true,scale=0.51]{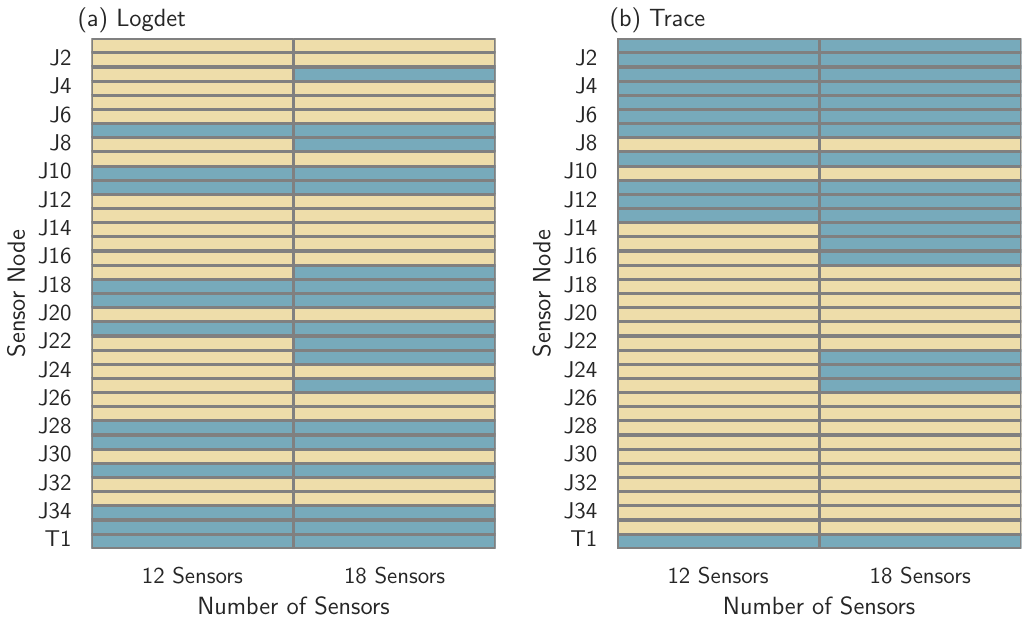}
	\vspace{-0.5cm}
	\caption{Optimal sensor configuration $\mc{S}^{*}$ obtained by solving $\mathbf{P1}$ using Algorithm~\ref{algorithm1} with $\kappa=3$ for Net2. The result in (a) is for the objective function $\mc{O}_{\mr{log\,det}}$, and in (b) for $\mc{O}_{\mr{trace}}$.}\label{fig:NET2-OSP}
\end{figure}

To further discuss this result, we first solve the SP problem while considering the three hydraulic patterns presented in Fig.~\ref{fig:demand2}. The optimal results from solving $\mathbf{P1}$ for Net2 under $\kappa \in \{1,2,3\}$ using Algorithm~\ref{algorithm1} for the two objective functions are presented in Fig.~\ref{fig:NET2-OSP}. The results for Net2 are consistent with those for Net1 in that they retain the submodularity of the solution. As mentioned above, this modularity allows for the framework’s use in operational settings that involve WQ networks where additional sensors may be required over time.

The results also show that the SP solved for the $\mc{O}_{\mr{log\,det}}$ observability measure still produces different results than the $\mc{O}_{\mr{trace}}$ measure. The reason is due to the distinct interpretation of the two measures as discussed in Section~\ref{sec:Obs-WQD}. That being said, the choice of measure for the SP objective function depends on specific operational needs, often defined by the system operator of the water network. From such operators perspective, the trace metric is suitable when the placement of WQ sensors is required to observe average WQ concentrations across the system. This means that certain WQ concentrations within branched pipes are often neglected, making it ideal for networks with more uniform flow patterns. As for the $\mc{O}_{\mr{log\,det}}$ measure, it ensures maximum coverage and estimability of WQ concentrations across all considered network components. This is ideal for networks that exhibit branching and under a higher hydraulic variability. This is illustrated in Fig.~\ref{fig:NET2-OSP}, where the network nodes selected for the $\mc{O}_{\mr{log\,det}}$ case are more dispersed and belong to at least one community structure within the network, thereby sensing the nodes within that branch or loop. For instance, the chosen sensor configuration $\mc{S}_2$, which includes $\{\mr{J3},\mr{J8}, \mr{J13},\mr{J18}, \mr{J21}, \mr{J28}\}$, consists of nodes that all belong to a specific branch or loop in the network. In contrast, the sensor configuration for that of $\mc{O}_{\mr{trace}}$ results in a more localized and centralized sensor configuration $\mc{S}_2$, which includes $\{\mr{J9},\mr{J11}, \mr{J12},\mr{J13}, \mr{T1}, \mr{J24}\}$. 

\subsection{Computational Complexity of the Proposed SP Framework}\label{sec:compt_time}
\begin{figure}[t]
	\includegraphics[keepaspectratio=true,scale=0.6]{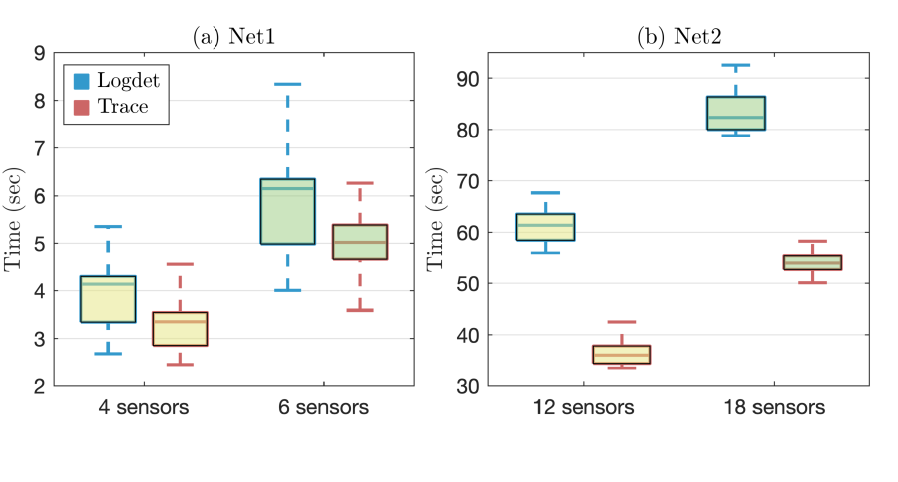}
	\vspace{-1cm}
	\caption{Computational time for solving $\mathbf{P1}$ for each network: (a) Net1, and (b) Net2.}
	\label{fig:OSP-TIME}
	\vspace{-0.2cm}
\end{figure}

The computational effort required to solve $\mathbf{P1}$ for (a) Net1 and (b) Net2, for each of the considered objective functions, is depicted in Fig.~\ref{fig:OSP-TIME}. For both networks, note that (i) the number of sensors chosen and (ii) the definition of the objective function affect the computational time of the submodular optimization problem~$\mathbf{P1}$. That being said, the greedy algorithm has a time complexity of $\mc{O}(|\mc{S}||\mc{N}|)$, which is significantly lower than exhaustive search approaches for integer programs, which have an exponential time complexity of $\mc{O}(2^{|\mc{N}|})$\cite{Calinescu2011}. This concurs with the results in Fig.~\ref{fig:OSP-TIME}, showing that the cardinality of the sensor set $\mc{S}$ affects the computational burden. 
For the case of Net1 under $\mc{O}_{\mr{log\,det}}$ objective function, the average time to solve $\mathbf{P1}$ is around $4$ secs, with the 25th and 75th percentiles at $[3.35, 4.3]$ secs for selecting a sensor configuration of $|\mc{S}_1| = 4$ sensors. This increases to an average of $5.73$ secs, with a percentile range of $[4.99, 6.34]$, for a sensor configuration of $|\mc{S}_2| = 6$ sensors. This variability is due to the computation of the observability Gramian~\eqref{eq:param_obs_gramian_kappa}, which depends on the WQ state-space vector $\hat{\vx}_0^{(\kappa)}$. The WQ concentrations vary according to the prevailing hydraulic conditions, causing $n_x$ to change between the considered cases. We note that the computational effort, as depicted for both networks under the $\mc{O}_{\mr{trace}}$ observability measure, is around $15\%$ to $40\%$ less exhaustive on average due to the simpler evaluation of the trace function. For the case of Net2, the computational effort increases to a mean of $58.32$ secs for $\mc{O}_{\mr{log\,det}}$ and $35.59$ secs for $\mc{O}_{\mr{trace}}$ for $|\mc{S}_1| = 12$. This complexity increases to $103.78$ secs and $54.29$ secs for $|\mc{S}_1| = 18$. This indeed is consistent with the aforementioned time complexity $\mc{O}(|\mc{S}||\mc{N}|)$ of the greedy algorithm. This concludes the case studies. 

\section{Conclusion, Paper Limitations, and Recommendations for Future Work}\label{sec:summary}
A robust optimal SP framework for multi-species WQ networks is proposed from a control-theoretic perspective. In contrast to other methods in the literature, our approach is applicable to nonlinear WQ models that represent chlorine and other contaminants, while also modeling their reactivity within the network. Accordingly, the SP framework is then posed as a submodular set maximization problem under a cardinality constraint. The framework considers maximizing observability-based measures that inform system operators on where to optimally allocate sensors within a network by considering different operational aspects. Furthermore, while accounting for the nonlinear dynamics of multi-species reactivity, the optimal sensor configuration obtained accounts for different WQ conditions and different hydraulic profiles that can exist in the daily operation of the WDN. A single optimal sensor configuration is obtained by solving the proposed problem under different hydraulic and WQ cases using a greedy algorithm. The resulting problem is less computationally challenging than exhaustive search approaches, thereby rendering the problem applicable to SP in larger networks. 

The paper's limitations and corresponding opportunities for future work are given as follows. 
\begin{itemize}[leftmargin=*]
	\item We do not study the state estimation of the WQ dynamics within the network. However, the proposed framework accounts for the ability to estimate states of WQ dynamics based on the results presented in Section~\ref{sec:Obs-WQD}. Furthermore, we do not account for partial knowledge of the hydraulic demand patterns. The framework can be extended to account for such cases by estimating \textit{a priori} the system hydraulics resulting from missing demand patterns. Additionally, although real-world hydraulic patterns and WQ reactants were not utilized, the theoretical developments of the proposed generalized SP framework allow for their inclusion in future implementations.
	\item Another aspect that we do not fully investigate is the high dimensionality of the state-space representation of the multi-species WQ dynamics, which can result from attaining stable and accurate numerical solutions under varying hydraulic settings. This is evident from the presented computational results, which vary based on the hydraulic and WQ case chosen. A plausible solution for this aspect would be to consider reduced-order nonlinear models that maintain the performance and robustness of the WQ dynamics while achieving a more scalable framework; see~\cite{Elsherif2024b}.
	\item Furthermore, the SP problem is posed and solved under a cardinality constraint that reflects the number of sensors deployed within the network. Additional operational constraints, such as those reflecting budget costs associated with the constraint \( |\mc{S}| = r \), or those targeting contamination detection and system resilience, can also be considered. In our recent work, we demonstrated that the SP problem attains a similar performance guarantee under a sensor budget constraint~\cite{Kazma2024a}.
	\item Finally, allocation of WQ sensors for contamination events can be considered within the presented framework by studying different submodular observability measures, such as the minimum eigenvalue. This enables the operator to allocate sensors in locations where minimal detection is available. To that end, additional observability-based measures, or a combination thereof, can be considered to help inform the operator on the optimal sensor configurations for different operational scenarios, such as investigating the WDN resilience and vulnerability.
\end{itemize}
Future research on this topic can investigate some or a combination of the above limitations of this present paper. 






\newpage
\onecolumn
\appendices
\section{Multi-Species Water Quality Models}\label{apndx:hydraulic_model} 
\begin{table*}[h!]
	\centering
	\caption{Multi-Species Water Quality Models for Different Network Components}\label{tab:HydModel}
	{\small	\begin{tabular}{|E|Q|W|K|}
			\hline
			& Model & Variables/Parameters & Model Description, Assumptions, and/or Limitations  \\
			\hline
			Reservoirs & $c_i^\mr{R}(k+1)=c_i^\mr{R}(k)$ & --- & Constant concentrations at reservoirs
			\\		
			\hline
			Pumps and Valves & {$c_i^{\mr{M}}(k+1) = c_l^{\mr{N}}(k+1),$ $\;\; c_j^{\mr{V}}(k+1) = c_l^{\mr{N}}(k+1)$} 
			& $c_l^{\mr{N}}(k+1)$ is the concentration at node $l$ upstream the pump/valve 
			\vspace{+0.15cm}
			& Transmission links with concentration equals the concentration of the node upstream		
			\\		
			\hline 
			Junctions & $c_i^\mr{J}(k)= \frac{\sum_{j \in L_{\mr{in}}} q_{\mr{in}}^{j}(k) c_\mr{in}^j(k)+q^\mr{B_\mr{J}}_i(k) c^\mr{B_\mr{J}}_i(k)}{q^{\mr{D}_\mr{J}}_i(k)+\sum_{l \in L_{\mr{out}}} q_{\mr{out}}^{l}(k)}$ 
			&\vspace{+0.15cm} $L_{\mr{in}}$ links flowing into the junction and $L_{\mr{out}}$ links extracting flow from the junction; $q_{\mr{in}}^{j}(k)$ and $q_{\mr{out}}^{k}(k)$ are the inflows and outflows from these links connected to the junction; $c_\mr{in}^j(k)$ is the concentration in the inflow solute;  $q^\mr{B_\mr{J}}_i(k)$ is the flow injected to the junction with concentration $c^\mr{B_\mr{J}}_i(k)$ by booster station if located; and $q^{\mr{D}_\mr{J}}_i(k)$ is demand  &  Complete and instantaneous mixing at junctions \\
			\hline
			Tanks & \vspace{+0.1cm}$\begin{aligned}
				&V_i^\mr{TK}(k+1) c_i^\mr{TK}(k+1)= V_i^\mr{TK}(k) c_i^\mr{TK}(k) \\
				&\quad\quad+\sum_{j \in L_{\mr{in}}} q^j_\mr{in}(k)c^j_\mr{in}(k) \Delta t_\mathrm{WQ}\\ 
				&\quad\quad+V^\mr{B_\mr{TK}}_i(k+1)c^\mr{B_\mr{TK}}_i(k+1) \\
				&\quad\quad- \sum_{l \in L_{\mr{out}}} q^l_\mr{out}(k)c_i^\mr{TK}(k) \Delta t_\mathrm{WQ}\\ 
				&\quad\quad+R^\mr{TK}_{\mr{MS}}(c_i^\mr{TK}(k)) V_i^\mr{TK}(k) \Delta t_\mathrm{WQ} 
			\end{aligned}$
			
			&  \vspace{+0.15cm}$V^\mr{B_\mr{TK}}_i(k+1)$ is the volume injected to the tank with concentration $c^\mr{B_\mr{TK}}_i(k+1)$ by booster station if located. $R^\mr{TK}_{\mr{MS}}(c^\mr{TK}_i(k))$ is the multi-species reaction rate in tanks & Complete instantaneous mixing of all inflows, outflows, and stored water following the continuously stirred tank reactor model \\
			\hline
			Pipes & \vspace{-1cm}$\begin{aligned}
				&	\text{First segment with upstream Junction $j$} \\
				&		c^\mr{P}_i(1,k+1) = (1-{\lambda}_i(k)) c^\mr{P}_i(1,k)\\ &\hspace{1.8cm}+{\lambda}_i(k) c^\mr{J}_j(k)\\ & \hspace{1.8cm} +R^\mr{P}_{\mr{MS}}(c^\mr{P}_i(s,k)) \Delta t_\mathrm{WQ}, \\
				&	\text{Any other segment} \\
				&  c^\mr{P}_i(s,k+ 1) = (1-{\lambda}_i(k)) c^\mr{P}_i(s,k)\\ &\hspace{1.8cm}+{\lambda}_i(k) c^\mr{P}_i(s-1,k)\\
				& \hspace{1.8cm} +R^\mr{P}_{\mr{MS}}(c^\mr{P}_i(s,k)) \Delta t_\mathrm{WQ},
			\end{aligned}$ & Pipe $i$ with length $L_{i}$ is split into a number of segments $s_i=\Big\lfloor \frac{L_i}{v_i(k) \Delta t_\mathrm{WQ}} \Big\rfloor$ of length $\Delta x_i= \frac{L_i}{s_i}$. $R^\mr{P}_{\mr{MS}}(c^\mr{P}_i(x,k))$ is the multi-species reaction rate in pipes. & The advection-reaction partial differential equation is discretized using upwind scheme. ${\lambda}_i(k)=\frac{v_i(k) \Delta t_\mathrm{WQ}}{\Delta x_i}$ is the Courant number and according to Courant-Friedrichs-Lewy condition (CFL), Courant number (CN) is maintained to be in the range of  $0<{\lambda}_i(k) \leq 1$ so that the scheme is stable.  \\
			\hline
			Multi-Species Dynamics & \vspace{-0.1cm}$\begin{aligned}
				&\hspace{-0.1cm} R^{\mr{P}}_{\mr{MS}}(c^\mr{P}_i(s,k))=-(\alpha_i^{\mr{P}_w}+\alpha_r \tilde{c}^\mr{P}_i(s,k)){c}^\mr{P}_i(s,k), \\
				& \hspace{-0.1cm}R^{\mr{P}}_{\mr{MS}}(\tilde{c}^\mr{P}_i(s,k))=-\alpha_r {c}^{\mr{P}}_i(s,k) \tilde{c}^\mr{P}_i(s,k),\\
				& \hspace{-0.1cm}R^{\mr{TK}}_{\mr{MS}}(c^\mr{TK}_i(k))=-(\alpha_b+\alpha_r \tilde{c}^\mr{TK}_i(k)){c}^\mr{TK}_i(k), \\
				& \hspace{-0.1cm}R^{\mr{TK}}_{\mr{MS}}(\tilde{c}^\mr{TK}_i(k))=-\alpha_r {c}^\mr{TK}_i(k) \tilde{c}^\mr{TK}_i(k)
			\end{aligned}$ &  $\alpha_r$ is the mutual reaction coefficient; $k_{b}$ is the bulk reaction rate constant; $\alpha_i^{\mr{P}_w} = \alpha_{b}+\frac{2\alpha_{w}\alpha_{f}}{r_{\mr{P}_i}(\alpha_{w}+\alpha_{f})},$ is the Pipe $i$ wall reaction coefficient; $\alpha_{w}$ is the wall reaction rate constant; $\alpha_{f}$ is the mass transfer coefficient between the bulk flow and the pipe wall; $r_{\mr{P}_i}$ is the pipe radius & First-order decay model, in addition to second-order reaction model between chlorine and a fictitious reactant. Only applicable in pipes and tanks.	\\
			\hline
			\hline
	\end{tabular}}
\end{table*}

\twocolumn
\section{Main Derivations}\label{apndx:proofs}
\subsection{Derivation of Gramian~\eqref{eq:param_obs_gramian}}\label{apndx:Gram}
To derive the parameterized observability Gramian~\eqref{eq:param_obs_gramian} from~\eqref{eq:obs_gramian_x0} of the multi-species WQ around a particular initial state $\hat{\vx}_0$ and with $\mc{S} \subseteq \mc{N}$ models~\eqref{eq:DT_dynamics_compact}, we start with the following. From  \eqref{eq:obs_jacobian_x0_param} and \eqref{eq:obs_gramian_x0}, it follows that for $\tilde{\vx} = \{\hat{\vx}_i\}^{N_s-1}_{i=0}$ we have 
\begin{align}\label{eq:proof1}
		\hspace{-0.2cm}\m{W}(\mc{S},\hat{\vx}_0) 
		&= {\left[\frac{\partial \tilde{\vx}}{\partial \hat{\vx}_0}\right]}^{\top}\bmat{\m I \otimes \m \Gamma \m C}^{\top}
		\bmat{\m I \otimes \m \Gamma \m C} \times  {\left[\frac{\partial \tilde{\vx}}{\partial \hat{\vx}_0}\right]}
\end{align}
Since the matrix multiplication in~\eqref{eq:proof1} is equivalent to an inner product, representing the summation of the matrix multiplication entries along the columns, and given that $\m \Gamma^2 = \m \Gamma$ and $\m C^\top \m \Gamma \m C$ is equivalent to $\left(\sum_{j=1}^{n_y} \gamma_j \m c_j^\top \m c_j \right)$, we can write the following
\begin{align*}
	\hspace{-2.5cm}\m{W}(\mc{S},\hat{\vx}_0) 
	&= \sum_{i=0}^{N_s-1}\left({\frac{\partial \hat{\vx}_i}{\partial \hat{\vx}_0}}\right)^{\hspace{-0.1cm}\top} \m C^\top \m \Gamma^2 \m C {\frac{\partial \hat{\vx}_i}{\partial \hat{\vx}_0}},\nonumber \\
	&= \sum_{i=0}^{N_s-1}\left({\frac{\partial \hat{\vx}_i}{\partial \hat{\vx}_0}}\right)^{\hspace{-0.1cm}\top}\left(\sum_{j=1}^{n_y} \gamma_j\m c_j^\top \m c_j \right) {\frac{\partial \hat{\vx}_i}{\partial \hat{\vx}_0}},\nonumber \\
	&= \sum_{i=0}^{N_s-1}\sum_{j=1}^{n_y}\gamma_j\left({\frac{\partial \hat{\vx}_i}{\partial \hat{\vx}_0}}\right)^{\hspace{-0.1cm}\top} \m c_j^\top \m c_j {\frac{\partial \hat{\vx}_i}{\partial \hat{\vx}_0}}, \label{eq:param_obs_gramian_proof}
	\\
	&= \sum_{j=1}^{n_y} \gamma_j \left(\sum_{i=0}^{N_s-1}\left({\frac{\partial \hat{\vx}_i}{\partial \hat{\vx}_0}}\right)^{\hspace{-0.1cm}\top} \hspace{-0.00cm}\m c_j^\top \m c_j {\frac{\partial \hat{\vx}_i}{\partial \hat{\vx}_0}}\right),\nonumber\\
	&= \sum_{j\in \mc{S}} \left(\sum_{i=0}^{N_s-1}
	\left(\dfrac{\partial \hat{\vx}_i}{\partial \hat{\vx}_0}\right)^{\hspace{-0.1cm}\top} 	\hspace{-0.075cm}\m c_j^\top \m c_j \dfrac{\partial \hat{\vx}_i}{\partial \hat{\vx}_0}\right),\nonumber
\end{align*}
where $ \sum_{j\in \mc{S}} $ is equivalent to the summation of the activated sensor nodes from parameterization $\m{\gamma}$ denoted as $\sum_{j=1}^{n_y} \gamma_j$.

\subsection{Derivation of Submodular Observability Measures}\label{apndx:Submod}
In this sequel we will analyze the modularity and submodularity properties of the proposed observability measures. We analyze the cases when the function $\mc{L}(\cdot)$ is $\mr{trace}$, and $\mr{log\,det}$. The following property provides support to the analysis on the modularity, submodularity, and monotonicity properties of the aforementioned measures.

Conic combinations, along with set restrictions and contractions, are submodularity-preserving operations~\cite{Bach2013b}. That being said, the submodularity of the original submodular functions is retained under a non-negative weighted sum. Let the set of submodular functions as be denoted as $\mc{L}_{1}, \mc{L}_{2}, \dots, \mc{L}_{\kappa}: 2^{\mc{N}}\rightarrow \mbb{R}$. Their conic combination can be written as follows
\begin{equation}
	\mc{O}(\mc{S})\hspace{-0.05cm}:=\hspace{-0.05cm}\sum_{\kappa=1}^{d}w_{\kappa}\mc{L}_{\kappa},
\end{equation}
such that $\mc{O}(\mc{S})$ retains the submodular set properties, given that $w_{\kappa} \geq 0 \;\forall \;\kappa$.

Now, consider the case when $\mc{L}_{\kappa} = \mr{trace}({\m W}^{(\kappa)}(\mc{S},\hat{\vx}_0))$; this results in the objective function $\mc{O}_{\mr{trace}}$. Then, for any $\mc{S} \subseteq \mc{N}$, observe that
\begin{align*}
		\mc{O}(\mc{S}) &= \frac{1}{d}\sum_{\kappa=1}^{d}\mr{trace}\left({\m{W}}^{(\kappa)}(\mc{S}, \hat{\vx}_0)\right), \\
	&= \frac{1}{d}\sum_{\kappa=1}^{d}\mr{trace}\left(\sum_{j\in \mc{S}} \left(\sum_{i=0}^{N_s-1}
	\left(\dfrac{\partial \hat{\vx}_i^{(\kappa)}}{\partial \hat{\vx}_0^{(\kappa)}}\right)^{\hspace{-0.1cm}\top} 	\hspace{-0.075cm}\m c_j^\top \m c_j \dfrac{\partial \hat{\vx}_i^{(\kappa)}}{\partial \hat{\vx}_0^{(\kappa)}}\right)\hspace{-0.2cm}\right), \\
	&= \sum_{j\in \mc{S}}  \left(\frac{1}{d}\sum_{\kappa=1}^{d}\mr{trace}\left(\sum_{i=0}^{N_s-1}
	\left(\dfrac{\partial \hat{\vx}_i^{(\kappa)}}{\partial \hat{\vx}_0^{(\kappa)}}\right)^{\hspace{-0.1cm}\top} 	\hspace{-0.075cm}\m c_j^\top \m c_j \dfrac{\partial \hat{\vx}_i^{(\kappa)}}{\partial \hat{\vx}_0^{(\kappa)}}\right)\hspace{-0.2cm}\right).
\end{align*}	

This shows that the $\mr{trace}(\cdot)$ under a conic combination is a linear matrix function and thus retains its modularity. A similar result holds true for the case when $\mc{L}_{\kappa} = \mr{log\;det}({\m W}^{(\kappa)}(\mc{S},\hat{\vx}_0))$; however, for brevity we do not include this in the derivation; see~\cite{Kazma2023f}.


\balance
\bibliographystyle{IEEEtran}
\bibliography{library.bib}

\end{document}